\documentstyle[
preprint,
prd,aps
,epsf
]{revtex}
\begin{document}
\noindent 
\hspace*{73ex} OCU-PHYS-182 \\ 
\hspace*{73.5ex} September 2001 
\begin{center} 
{\LARGE Rates of neutrino conversion and decay} \\ 
{\LARGE in hot and dense QED plasma} 
\end{center} 
\begin{center} 
{\large Asida N.}\footnote{Electronic address: 
asida@yukawa.kyoto-u.ac.jp} \\ 
{\it Hukuyama City Junior College for Women, Hirosima-ken 
Hukuyama-si} \\ 
{\it Kita-Honzyoo 720-0074, JAPAN} 
\end{center} 
\begin{center} 
{\large A. Ni\'{e}gawa}\footnote{Electronic address: 
niegawa@sci.osaka-cu.ac.jp} {\large and H. 
Ozaki}\footnote{Electronic address: hozaki@sci.osaka-cu.ac.jp} \\ 
{\it Department of Physics, Osaka City University, 
Sumiyoshi-ku, Osaka 558-8585, JAPAN} 
\end{center} 
\begin{center} 
{\large M. Kubota} \\ 
{\it Hitachi Software Engineering Co., Ltd., Naka-ku, 
Yokohama 231-0015, JAPAN} 
\end{center} 
\hspace*{4ex} 
\begin{center} 
{\large Abstract} 
\end{center} 
Using a real-time formalism of equilibrium and nonequilibrium 
quantum-field theory, we derive the reaction-rate formula for 
neutrino-conversion ($\nu \to \nu'$) process and $\nu \bar{\nu}'$ 
annihilation process, which take place in a hot and dense QED 
plasma with background (anti)neutrinos out of equilibrium. Also 
derived is the formula for the inverse processes to the above ones. 
Using the hard-thermal-loop resummation scheme, we include the 
contribution from the coherent processes. The decay/production of a 
neutrino causes an evolution of its spatial distribution. A scheme 
for dealing with this evolution is presented. For the case of 
isotropic neutrino distribution, numerical computation is carried 
out for the parameter region of type-II super-nova explosion. 
Defferential reaction rate exhibits characteristic peak structure, 
which comes from the coherent processes. The contribution from the 
above processes to the decay or damping rate of a parent neutrino 
$\nu$ is also studied. 

\hspace*{1ex}

11.10.Wx, 13.10.+q, 13.15.+g, 13.35.Hb

\hspace*{1ex}

\hspace*{1ex}
\narrowtext 
\section{Introduction} 
\def\theequation{\mbox{\arabic{equation}}}
For the past two decades, properties of neutrinos in background 
media have attracted much interest (see, e.g., \cite{pecc}). 
Interactions of neutrinos with a thermal background cause a change 
in the properties of neutrinos. The dispersion relation is the 
quantity that describes this change. N\"otzold and Raffelt 
\cite{disp} were the first who comprehensively analyzed the 
dispersion relation of a neutrino in a thermal background, where, 
among others, the damping rate of an electron neutrino 
is computed, the rate which is related to the mean-free path and to 
the refractive index. Computation is performed by neglecting Pauli 
blocking effects and using the bare dispersion relation for 
participating electrons. Radiative decay of a massive neutrino has 
been analyzed in \cite{olivo}. 

It is by now well known \cite{le-b,pis} that, in hot and/or dense 
QED, the thermal propagators of a soft photon\footnote{A soft 
particle is the particle that carries soft momentum $Q^\mu$ 
($|Q^\mu| = O (e \sqrt{T^2 + \mu^2})$). Here $- e$ is the electron 
charge and $T$ $(\mu)$ denotes the temperature (chemical 
potential).} and a soft electron (positron) are drastically changed 
from those of respective bare counterparts. The salient feature is 
the appearance of the imaginary part in space-like-momentum region, 
which comes from Landau damping mechanism. The dispersion relations 
for soft photon and electron are also largely changed. An effective 
or improved perturbation theory, called hard-thermal-loop (HTL) 
resummation scheme \cite{le-b,pis}, in which the above-mentioned 
effects are taken into account, is established just after the work 
\cite{disp}. 

Recently, in relation to the possible neutrino oscillation, 
neu\-tri\-no--con\-ver\-sion processes have attracted much interest. 
For instance, in \cite{ayala}, the rate of neutrino conversion 
$\nu_L \to \nu_R$ in a hot and dense QED (or 
e\-lec\-tron--pos\-i\-tron--pho\-ton) plasma is computed using the 
above-mentioned effective perturbation theory. The responsible 
interaction is the magnetic dipole neutrino-photon interaction. 
$\nu_L$ is assumed to be thermally distributed in the QED plasma. 

In this paper, we deal with a neutrino-conversion ($\nu \to \nu'$) 
process and a $\nu \bar{\nu}'$ annihilation process, which take 
place in a hot and dense QED plasma. We also deal with inverse 
processes to them. The background (anti)neutrinos are out of 
equilibrium. In Sec.~II, we derive the reaction-rate formula for 
these processes, on the basis of the nonequilibrium quantum-field 
theory (outlined in Appendix A) supplemented with the effective 
perturbation theory of hot and dense QED. Also presented is a 
procedure of determining spacetime evolution of (anti)neutrino 
distribution function. In Sec.~III, for the purpose of illustration, 
numerical computation is carried out for the differential reaction 
rate for the case of isotropic neutrino distribution. We are 
interested in the temperature and baryon-number density regions of 
type-II supernova explosion (cf. \cite{le-b,loulet}); $m_e << T, \; 
\mu << m_{ion}$. As in \cite{olivo,ayala}, we neglect the effect of 
the ions. The contribution from the coherent processes exhibits a 
characteristic peak structure in energy distribution of a \lq\lq 
decay neutrino'' --- $\nu'$ for the neutrino-conversion process and 
$\bar{\nu}'$ for the $\nu \bar{\nu}'$ annihilation process. We then 
study the contribution to the damping rate of $\nu$. Numerical 
computation is carried out for the case where no neutrino exists in 
background. Section IV is devoted to discussions. In Appendix A, we 
present a formalism for dealing with neutrinos in out of equilibrium 
system, which is used in the main text. In Appendix B, we give the 
form for the self-energy-part-resummed photon propagator. Finally, 
in Appendix C, we compute the necessary quantities for the present 
study. 
\section{Reaction-rate formula and damping rate of a neutrino} 
\subsection{Neutrino-conversion process} 
We deal with the system that consists of a hot and dense QED plasma 
and neutrinos. We assume that the QED plasma is in thermal and 
chemical equilibrium, while the background neutrinos are not. Then, 
the whole system is out of equilibrium. A neutrino-conversion 
process of our concern is 
\begin{equation} 
\nu (K) + \mbox{QED plasma} \to \nu' (K') + 
\mbox{anything} . 
\label{process1} 
\end{equation} 
We assume that $\nu$ is massive, left-handed neutrino (with mass 
$m$), $\nu'$ is massless and left-handed. The four-momenta $K$ and 
$K'$ are $K = (E, {\bf k})$ with $E = \sqrt{k^2 + m^2}$ and $K' = 
(k', {\bf k}')$, respectively. The QED plasma is assumed to be at 
rest. The total reaction rate for the process (\ref{process1}) 
contributes to the damping or \lq\lq decay'' rate $\Gamma_d$ of a 
parent neutrino $\nu$. $\Gamma_d$ also receives a contribution from 
the relative process to (\ref{process1}), 
\begin{equation} 
\nu (K) + \bar{\nu}' (- K') + \mbox{QED plasma} \to 
\mbox{anything} , 
\label{process2} 
\end{equation} 
where $\bar{\nu}'$ is an antiparticle of $\nu'$ and $- K' = (k', - 
{\bf k}')$. 

The region for various parameters of our interest is 
\begin{equation} 
T, \, \mu, E, \, k', \, m_e << M_W \;\;\;\; \mbox{and} \;\;\;\; m < 
2 m_e . 
\label{mas} 
\end{equation} 
Here $M_W$ $(m_e)$ is the mass of $W$ boson (electron). We expect 
that, in the region (\ref{mas}), the result is insensitive to $m$ 
($< m_e$). As a matter of fact, we have confirmed that, at least 
in the range $0 \leq m \leq 100$ eV/c$^2$ \cite{AG}, no visible 
change is detected in the results to be given in Sec.~III. Thus, 
we set $m = 0$ throughout in the sequel, so that $K = (k, {\bf 
k})$. In the region (\ref{mas}), we may use the effective 
Lagrangian, which, after Fierz transformation, reads 
\[ 
{\cal L}_{\mbox{\scriptsize{eff}}} = - 2 \sqrt{2} G \left[ \left( 
\bar{\underline{\nu}}' \gamma^\mu L \underline{\nu} \right) \left( 
\bar{\underline{e}} \gamma_\mu L \underline{e} \right) + \mbox{h.c.} 
\right] . 
\] 
Here $L \equiv (1 - \gamma_5) / 2$, $G$ is Fermi's constant, and 
underlined fields stand for the fields {\em in the \lq\lq 
weak-interaction basis''}. 

For the QED sector, we employ the real-time formalism of equilibrium 
QED \cite{le-b,lan} and, for the neutrino sector, we use the 
nonequilibrium real-time formalism of quantum-field theory 
\cite{chou}. $\Gamma_d$ is written as \cite{chou,nie} 
\begin{eqnarray} 
\Gamma_d & = & |U^*_{e \nu} \, U_{e \nu'}|^2 \, \tilde{\Gamma}_d , 
\label{rate00} \\ 
\tilde{\Gamma}_d & = & - \frac{i}{2 k} \, \mbox{Tr} \left[ L 
{K\kern-0.1em\raise0.3ex\llap{/}\kern0.15em\relax} \Sigma_{2 
1}^{(\nu)} (X; K) \right] , 
\label{sora} \\ 
& = & \int_{- \infty}^{+ \infty} d q_0 \, 
\frac{d \tilde{\Gamma}_d}{d q_0} , 
\label{gam} 
\end{eqnarray} 
where $U$ is the lepton mixing matrix and $\Sigma_{2 1}^{(\nu)}$ is 
the $(2 1)$-element of the $2 \times 2$ matrix self-energy part 
$\hat{\Sigma}_\nu (X; K)$ of nonequilibrium $\nu$. (Here and in the 
following, \lq\lq $\,\, \hat{} \,\,$'' denotes $2 \times 2$ matrix 
in a \lq\lq type-space'' \cite{le-b,lan,chou,nie}.) $X$ stands for 
the spacetime coordinates of the center of the region, where the 
reaction takes place, the definition of which is given shortly. The 
$O (G^2)$ contribution to $\tilde{\Gamma}_d$, 
$\tilde{\Gamma}_d^{(1)}$, reads \cite{chou,nie} 
\begin{eqnarray} 
\tilde{\Gamma}_d^{(1)} & = & - \frac{1}{2 k} \int \frac{d^{\, 4} 
Q}{(2 \pi)^4} \Pi_{2 1}^{(W) \alpha \beta} (Q) \nonumber \\ 
&& \mbox{\hspace*{6.5ex}} \times \mbox{Tr} \left[ L 
{K\kern-0.1em\raise0.3ex\llap{/}\kern0.15em\relax} \gamma_\alpha 
S^{(\nu')}_{2 1} (X; K - Q) \gamma_\beta L \right] , 
\label{W-pol} 
\end{eqnarray} 
where $\Pi_{2 1}^{(W) \alpha \beta}$ is the $(2 1)$ element of the 
one-loop matrix function $\hat{\Pi}_W^{\alpha \beta}$ whose ($i j$) 
element is 
\begin{eqnarray} 
\Pi_{i j}^{(W) \alpha \beta} (Q) & = & 8 i G^2 (-)^{i + j} 
\int \frac{d^{\, 4} P}{(2 \pi)^4} \, \mbox{Tr} \left[ S_{j i}^{(e)} 
(P - Q) \right. \nonumber \\ 
&& \left. \mbox{\hspace*{16ex}} \times \gamma^\alpha L S_{i j}^{(e)} 
(P) \, \gamma^\beta L \right] \nonumber \\ 
&& \mbox{\hspace*{25ex}} (i, j = 1, 2) . 
\label{W-pol1} 
\end{eqnarray} 
The diagrammatic representation of $\tilde{\Gamma}_d^{(1)}$ is given 
in Fig.~1. In Eq.~(\ref{W-pol1}), no summation is taken over $i$ and 
$j$. In Eq.~(\ref{W-pol}), $S^{(\nu')}_{2 1} (X; P)$ is the $(2 1)$ 
element of the matrix propagator $\hat{S}_{\nu'} (X; P)$ of out of 
equilibrium $\nu'$, which is a Wigner transform of its 
configuration-space counterpart, $\hat{S}_{\nu '} (x, y)$: 
\[ 
\hat{S}_{\nu '} (x, y) = \int \frac{d^{\, 4} P}{(2 \pi)^4} e^{- i P 
\cdot (x - y)} \hat{S}_{\nu'} (X; P) , 
\] 
where $X \equiv (x + y) / 2$. $S_{i j}^{(e)}$ in Eq.~(\ref{W-pol1}) 
is the $(i j)$ element of an equilibrium electron propagator matrix 
$\hat{S}_e$. 
\subsubsection*{Neutrino sector} 
From now on, we restrict our concern to the systems, in which the 
distributions of (anti)neutrinos are of quasi-uniform near 
equilibrium or of quasistationary. For such systems, the gradient 
approximation in the derivative expansion is sensible \cite{chou}. A 
brief derivation of $\hat{S}_{\nu'} (X; P)$ to the gradient 
approximation is given in Appendix A: $\hat{S}_{\nu'} (X; P) = 
\hat{S}^{(0)}_{\nu'} (X; P) + \hat{S}^{(1)}_{\nu'} (X; P)$ with 
$\hat{S}^{(0)}_{\nu'}$ the leading term and $\hat{S}^{(1)}_{\nu'}$ 
the nonleading or gradient term. $\hat{S}^{(0)}_{\nu'} (X; P)$ reads 
\begin{eqnarray} 
\hat{S}_{\nu'}^{(0)} (X; P) & = & L 
P\kern-0.1em\raise0.3ex\llap{/}\kern0.15em\relax 
\hat{\cal S}^{(0)}_{\nu'} (X; P) , \nonumber \\ 
\hat{\cal S}^{(0)}_{\nu'} (X; P) & = & \left( 
\begin{array}{cc} 
\Delta_R (P) & \; 0 \\ 
\Delta_R (P) - \Delta_A (P) & \; - \Delta_A (P)
\end{array} 
\right) \nonumber \\ 
&& + 2 \pi i \epsilon (p_0) f_{\nu'} (X; P) \delta (P^2) 
\hat{A}_+ , 
\label{zero} 
\end{eqnarray} 
where 
\begin{eqnarray} 
\Delta_{R (A)} (P) & = & \frac{1}{P^2 \pm i \epsilon (p_0) 0^+} , 
\label{ra} 
\\ 
f_{\nu'} (X; P) & = & \theta (p_0) N_{\nu'} (X; p_0, {\bf p}) 
\nonumber \\ 
&& + \theta (- p_0) \left[ 1 - N_{\bar{\nu}'} (X; |p_0|, - {\bf p}) 
\right] , 
\label{efu} 
\\ 
\hat{A}_\pm & = & \left( 
\begin{array}{cc} 
1 & \; \pm 1 \\ 
\pm 1 & \; 1 
\end{array} 
\right) . 
\label{Adesu} 
\end{eqnarray} 
Here $N_{\nu'} (X; p_0, {\bf p})$ [$N_{\bar{\nu}'} (X; |p_0|, - {\bf 
p})$] is the number-density function of $\nu'$ [$\bar{\nu}'$] at the 
space\-time point $X$ with \lq\lq energy'' $|p_0|$ and momentum 
${\bf p}$ [$- {\bf p}$]. The non\-leading term $\hat{S}^{(1)}_{\nu'} 
(X; P)$ is given in Eq.~(\ref{nupro2}) with $\nu'$ for $\nu$. In our 
formalism outlined in Appendix A, $\hat{S}^{(1)}_{\nu'}$ is $O 
(G^2)$ smaller than $\hat{S}^{(0)}_{\nu'}$. This is because 
$\hat{S}^{(1)}_{\nu'}$ includes $\partial_{X_\mu} f_{\nu'} (X; P)$, 
which, as seen from Eq.~(\ref{kj}) (with $\nu'$ for $\nu$), is 
proportional to $G^2$ since so is $F_{\nu'}$. Then, for $S_{2 
1}^{(\nu')}$ in Eq.~(\ref{W-pol}), we substitute the ($2 1$) 
element of the leading term $\hat{S}^{(0)}_{\nu'}$ 
(Eq.~(\ref{zero})). 
\subsubsection*{Electron sector} 
The electron propagator matrix $\hat{S}_e (P)$, whose elements are 
in Eq.~(\ref{W-pol1}), is of the form 
\begin{equation}
\hat{S}_e (P) = P\kern-0.1em\raise0.3ex\llap{/}\kern0.15em\relax 
\hat{\cal S}_e (P) . 
\label{esu} 
\end{equation} 
Here $\hat{\cal S}_e (P)$ is given by the right-hand side (RHS) of 
Eq.~(\ref{zero}) provided that $f_{\nu'} (X; P)$ is replaced by an 
equilibrium distribution function 
\begin{eqnarray}
f_e (p_0) & = & \theta (p_0) N_e (p_0) + \theta (- p_0) [ 1 - 
N_{e^+} (p_0) ] , \nonumber \\ 
N_e (p_0) & = & \frac{1}{e^{(p_0 - \mu) / T} + 1} , \nonumber \\ 
N_{e^+} (p_0) & = & \frac{1}{e^{(|p_0| + \mu) / T} + 1} . 
\label{bunp} 
\end{eqnarray} 

We decompose $\Pi_{i j}^{(W) \alpha \beta} (Q)$, Eq.~(\ref{W-pol1}), 
into transverse $(T)$, longitudinal $(L)$, and vector--axial-vector 
interference $(V\!A)$ parts. $T$- and $L$- parts are proportional to 
the thermal self-energy part of a photon in hot and dense QCD. Thus, 
we may write 
\begin{eqnarray} 
\Pi_{i j}^{(W) \alpha \beta} (Q) & = & 4 \frac{G^2}{e^2} \left[ 
{\cal P}_T^{\alpha \beta} (\hat{\bf q}) \, \Pi_{i j}^{(T)} (Q) + 
{\cal P}_L^{\alpha \beta} (Q) \, \Pi_{i j}^{(L)} (Q) \right. 
\nonumber \\ 
&& \left. - i \epsilon^{\alpha \beta \rho 0} Q_\rho \Pi_{i j}^{(V\! 
A)} (Q) \right] , 
\label{decomp2} \\ 
\Pi_{i j}^{(V\! A)} (Q) & = & 2 i e^2 \frac{Q^2}{q^2} \int 
\frac{d^{\, 4} P}{(2 \pi)^4} (q_0 - 2 p_0) \nonumber \\ 
&& \times {\cal S}_{j i}^{(e)} (P - Q) {\cal S}_{i 
j}^{(e)} (P) , \nonumber 
\end{eqnarray} 
where $\epsilon^{\mu \nu \rho \sigma}$ is a fully anti-symmetric 
pseudo tensor with $\epsilon^{0 1 2 3} = 1$, ${\cal S}_{i 
j}^{(e)}$ $(i, j = 1, 2)$ is the $(i j)$ element of $\hat{\cal 
S}_e$ in Eq.~(\ref{esu}), and ${\cal P}_{T (L)}^{\alpha \beta}$ is 
the standard projection operator onto transverse (longitudinal) 
mode, 
\begin{eqnarray}
{\cal P}_T^{\alpha \beta} (\hat{\bf q}) & = & - \sum_{i, j = 1}^3 
g^{\alpha i} g^{\beta j} (\delta^{i j} - \hat{q}^i \hat{q}^j) 
\;\;\;\;\;\;\;\; (\hat{q}^i \equiv q^i / q) , 
\nonumber \\ 
{\cal P}_L^{\alpha \beta} (Q) & = & g^{\alpha \beta} - 
\frac{Q^\alpha Q^\beta}{Q^2} - {\cal P}_T^{\alpha \beta} (\hat{\bf 
q}) . 
\label{proj} 
\end{eqnarray} 
$\Pi_{i j}^{(T / L)} (Q)$ in Eq.~(\ref{decomp2}) is the $(i j)$ 
element of the $T / L$-component of the thermal self-energy part of 
a photon.  

$\Pi_{2 1}^{(S)} (Q)$ $(S = T, L, V\!A)$ are related to 
\cite{le-b,lan} the so-called Feynman self-energy part 
\begin{eqnarray} 
\Pi_F^{(S)} (Q) & \equiv & \Pi_{1 1}^{(S)} (Q) 
+ \theta (q_0) \Pi_{1 2}^{(S)} (Q) + \theta (- q_0) 
\Pi_{2 1}^{(S)} (Q) \nonumber \\ 
&& \mbox{\hspace*{20ex}} (S = T, L, V\!A) , 
\label{PiF} 
\end{eqnarray} 
through 
\begin{eqnarray} 
i \Pi_{2 1 (1 2)}^{(S)} (Q) & = & 2 [\theta (\pm q_0) + n_B (|q_0|)] 
\, \mbox{Im} \Pi_F^{(S)} (Q) \nonumber \\ 
&& \mbox{\hspace*{15ex}} (S = T, L, V\!A) . 
\label{kan} 
\end{eqnarray} 
Here $n_B (x) = 1 / (e^{x / T} - 1)$ and \lq \lq $\mbox{Im}$'' means 
to take the imaginary part with Feynman prescription. 
In Appendix C, $\Pi_F^{(S)}$ $(S = T, L, V\!A)$ is computed within 
the approximation $m_e = 0$, which is a good approximation for a 
plasma with high temperature and/or density, $m_e << T, \mu$. 
In Sec.~III, we discuss to what extent the approximation $m_e = 0$ 
is good one. It is worth mentioning that $\Pi_F^{(V\!A)} (Q)$ 
vanishes for vanishing chemical potential, $\mu = 0$. 
\subsubsection*{$O (G^2)$ decay-rate formula} 
As seen from Eq.~(\ref{zero}), $S_{2 1}^{(\nu')} (X; K - Q)$ in 
Eq.~(\ref{W-pol}) involves $f_{\nu'} (X; K - Q)$. Let us 
write 
\begin{eqnarray*} 
f_{\nu'} (X; K - Q) & = & f_{\nu'} (X; k - q_0, {\bf k} - {\bf 
q}_{\parallel}, - {\bf q}_\perp) , \nonumber \\ 
{\bf q}_{\parallel} & \equiv & \frac{1}{k^2} ({\bf q} \cdot {\bf k} 
) {\bf k} , \;\;\;\;\;\; {\bf q}_\perp \equiv {\bf q} - {\bf 
q}_{\parallel} . 
\end{eqnarray*} 
Then we define $\bar{f}_{\nu'}$ through 
\begin{eqnarray} 
&& \int \frac{d^{\, 4} Q}{2 \pi} f_{\nu'} (X; K - Q) {\cal G} 
(q_0, {\bf q}_\parallel) \nonumber \\ 
&& \mbox{\hspace*{3ex}} = \int d q_0 d q \frac{d ({\bf q} \cdot {\bf 
k})}{q k} \, q^2 \bar{f}_{\nu'} \left( X; k - q_0, {\bf k} - {\bf 
q}_{\parallel} \right) {\cal G} (q_0, {\bf q}_\parallel) \nonumber 
\\ 
\label{iyaya} 
\end{eqnarray} 
with ${\cal G}$ any function of $q_0$ and ${\bf q}_\parallel$. 

Using the formulae displayed above, we obtain, after straightforward 
manipulation, for $d \tilde{\Gamma}_d^{(1)} / d q_0$ to $O(G^2)$ 
(see Eqs.~(\ref{gam}) and (\ref{W-pol})), 
\widetext 
\begin{eqnarray} 
\frac{d \tilde{\Gamma}_d^{(1)}}{d q_0} & = & \frac{G^2}{\pi^2} 
\frac{1}{e^2} \frac{1}{k^2} \left[ \theta (q_0) + n_B (|q_0|) 
\right] \epsilon (k - q_0) \nonumber \\ 
&& \times \int_{\cal R} d q \, q \, \left[ 1 - \bar{f}_{\nu'} \left( 
X; k - q_0, \left( 1 - \frac{q_0}{k} + \frac{Q^2}{2 k^2} \right) 
{\bf k} \right) \right] G^{(1)} (Q) , 
\label{1221} 
\end{eqnarray} 
\narrowtext

\noindent where 
\begin{eqnarray} 
G^{(1)} (Q) & = & \mbox{Im} \, \left[ H_T (Q) \Pi_F^{(T)} (Q) 
\right. \nonumber \\ 
&& \left. + H_L 
(Q) \Pi_F^{(L)} (Q) - H_{V\! A} (Q) \, \Pi_F^{(V\! A)} (Q) \right] , 
\nonumber \\ 
\label{G1} 
\\ 
H_T (Q) & = & Q^2 - 2 k^2 + \frac{2}{q^2} \left( k q_0 - 
\frac{Q^2}{2} \right)^2 , \nonumber \\ 
H_L (Q) & = & 2 k^2 - \frac{2}{q^2} \left( k q_0 - \frac{Q^2}{2} 
\right)^2 , \nonumber \\ 
H_{V\! A} (Q) & = & \left( q_0 - 2 k \right) Q^2 . \nonumber 
\end{eqnarray} 
The integration region ${\cal R}$ in Eq.~(\ref{1221}) is defined as 
(see Fig.~2) 
\begin{eqnarray*} 
{\cal R} & = & {\cal R}_1 \cup {\cal R}_2 , \nonumber \\ 
{\cal R}_1 & : & |q_0| \leq q \leq 2k - q_0 , 
\nonumber \\ 
{\cal R}_2 & : & |2k - q_0| \leq q \leq q_0 . 
\end{eqnarray*} 
In the region ${\cal R}_1$ [${\cal R}_2$], $k_0' = k - q_0 \geq 0$ 
[$k_0' < 0$], and then ${\cal R}_1$ [${\cal R}_2$] is the 
kinematically allowed region of the reaction (\ref{process1}) 
[(\ref{process2})]. At first sight, at $q_0 = 0$, 
Eq.~(\ref{1221}) seems to diverge at $q = 0$. Inspection of the 
formulae in Appendix C tells us, however, that this is not the case. 
\subsection{Neutrino-production process} 
We now turn to the inverse processes to (\ref{process1}) and to 
(\ref{process2}): 
\begin{eqnarray} 
&& \nu' (K') + \mbox{QED plasma} \to \nu (K) + 
\mbox{anything} , 
\label{pro1} \\ 
&& \mbox{QED plasma} \to \nu (K) + \bar{\nu}' (- K') 
+ \mbox{anything} , 
\label{pro2} 
\end{eqnarray} 
where $K' = (k', {\bf k}')$, $K = (k, {\bf k})$, and $- K' = (k', - 
{\bf k})$. The process (\ref{pro1}) is a production process of $\nu$ 
due to the reaction of $\nu'$ with constituents of the QED plasma 
and the process (\ref{pro2}) is a $\nu \bar{\nu}'$ production 
process. The reaction-rate formula for these processes is written as 
\begin{eqnarray} 
\Gamma_p & = & |U^*_{e \nu} \, U_{e \nu'}|^2 \, \tilde{\Gamma}_p , 
\label{rate0} \\ 
\tilde{\Gamma}_p & = & \frac{i}{2 k} \, \mbox{Tr} \left[ L 
{K\kern-0.1em\raise0.3ex\llap{/}\kern0.15em\relax} \Sigma_{1 
2}^{(\nu)} (X; K) \right] , 
\label{mori} \\ 
& = & \int_{- \infty}^{+ \infty} d q_0 \, 
\frac{d \tilde{\Gamma}_p}{d q_0} . 
\label{rate1} 
\end{eqnarray} 
In a similar manner as above, we obtain for the $O (G^2)$ 
contribution, 
\widetext 
\begin{eqnarray} 
\frac{d \tilde{\Gamma}_p^{(1)}}{d q_0} & = & - \frac{G^2}{\pi^2} 
\frac{1}{e^2} \frac{1}{k^2} \left[ \theta (- q_0) + n_B (|q_0|) 
\right] \epsilon (q_0 - k) \nonumber \\ 
&& \times \int_{\cal R} d q \, q \, \bar{f}_{\nu'} \left( 
X; k - q_0, \left( 1 - \frac{q_0}{k} + \frac{Q^2}{2 k^2} \right) 
{\bf k} \right) G^{(1)} 
(Q) , 
\label{prod} 
\end{eqnarray} 
\narrowtext 

\noindent 
where $\bar{f}_{\nu'}$ is as in Eq.~(\ref{iyaya}) and $G^{(1)} (Q)$ 
is as in Eq.~(\ref{G1}). The diagram for $\tilde{\Gamma}_p^{(1)}$ is 
the same as Fig.~1, provided that the two types of vertices are 
interchanged, $1 \leftrightarrow 2$. 
\subsection{Contributions from a set of coherent processes} 
According to the HTL-resummation scheme \cite{pis,le-b}, the 
integration region in Eqs.~(\ref{gam}), (\ref{1221}), (\ref{rate1}), 
and (\ref{prod}) should be divided into hard-$Q$ region ($|Q^\mu| 
= O ( \sqrt{T^2 + \mu^2})$) and the soft-$Q$ region ($|Q^\mu| = O 
(e \sqrt{T^2 + \mu^2})$). 

\noindent {\em Hard-$Q$ region}: For $\mbox{Im} \Pi_F^{(S)} (Q)$ $(S 
= T, L, V\!A)$, expressions given in Appendix C are used. \\ 
\noindent {\em Soft-$Q$ region}: 
Observing the formulae in Appendix C, we see that, for $e << 1$, 
\begin{eqnarray} 
H_{T / L} (Q) \, \mbox{Im} \Pi_F^{(T / L)} (Q) & \simeq & H_{T 
/ L} (Q) \, \mbox{Im} F_{T / L} (Q) \nonumber \\ 
& >> & H_{V \!A} (Q) \, \Pi_{2 1}^{(V A)} (Q) , 
\label{kinnji} 
\end{eqnarray} 
where $\mbox{Im} F_{T / L} (Q)$ is as in Eq.~(\ref{dai}) with 
Eqs.~(\ref{15}) and (\ref{Im-pi}) in Appendix C. In the soft-$Q$ 
region, there is an additional contribution: An inverse 
HTL-resummed photon propagator (cf. Eq.~(\ref{effe1})) $\left( 
\displaystyle{ \raisebox{0.6ex}{\scriptsize{*}}} \Delta_F^{(T / L)} 
(Q) \right)^{- 1} ( = Q^2 - \Pi_F^{(T / L)} (Q) \simeq Q^2 - F_{T 
/ L} (Q))$ is of the same order of magnitude as $\Pi_F^{(T / L)} 
(Q)$ $(\simeq F_{T / L} (Q))$. Thus, the diagram for 
$\tilde{\Gamma}_d$ as shown in Fig.~3 yields an equally 
important contribution. 

The characteristic scale of the hard region is \cite{pis,le-b} 
$\sqrt{T^2 + \mu^2}$, and that of the soft region is $e \sqrt{T^2 + 
\mu^2}$. As a matter of fact, since $e \simeq 0.30$, the hard region 
and the soft region are not {\em sharply separated}. Taking this 
fact into account, we compute the contribution from Fig.~3, 
$\Gamma^{(2)}$, without using the HTL-approximation $(e << 1)$. The 
contribution is given by Eq.~(\ref{W-pol}) with the replacement 
\cite{le-b,lan}, 
\begin{eqnarray} 
\Pi^{(W) \alpha \beta}_{2 1} (Q) & \to & - \frac{e^2}{8 G^2} 
\sum_{i, \, j = 1}^2 \Pi_{2 i}^{(W) \alpha \mu} (Q) \nonumber \\ 
&& \times \left( 
\displaystyle{\raisebox{0.6ex}{\scriptsize{*}}} \! \Delta_{i j} (Q) 
\right)_{\mu \nu} \Pi_{j 1}^{(W) \nu \beta} (Q) , 
\label{kiso} 
\end{eqnarray} 
where $\Pi$'s are as in Eq.~(\ref{W-pol1}) and 
$\displaystyle{\raisebox{0.6ex}{\scriptsize{*}}} \! \Delta$ is as in 
Appendix B. Straightforward computation yields 
\widetext 
\begin{eqnarray} 
\frac{d \tilde{\Gamma}^{(2)}_d}{d q_0} & = & \frac{G^2}{\pi^2} 
\frac{1}{e^2} \frac{1}{k^2} \left[ \theta (q_0) + n_B (|q_0|) 
\right] \epsilon (k - q_0) \nonumber \\ 
&& \times \int_{\cal R} d q \, q \, \left[ 1 - \bar{f}_{\nu'} 
\left( X; k - q_0, \left( 1 - \frac{q_0}{k} + \frac{Q^2}{2 k^2} 
\right) {\bf k} \right) \right] G^{(2)} (Q) , 
\label{sof} 
\end{eqnarray} 
where 
\begin{eqnarray}
G^{(2)} (Q) & = & \frac{1}{2} \, \mbox{Im} \left[ \sum_{P = T, \, 
L} H_P (Q) \frac{\left( \Pi_F^{(P)} (Q) \right)^2}{Q^2 - 
\Pi_F^{(P)} (Q)} \right. 
\nonumber \\ 
&& \left. + H_T (Q) q^2 \, \frac{\left( \Pi_F^{(V\! A)} (Q) 
\right)^2}{Q^2 - \Pi_F^{(T)} (Q)} - 2 H_{V \! A} (Q) \, 
\frac{\Pi_F^{(T)} (Q) \Pi_F^{(V\! A)} (Q)}{Q^2 - \Pi_F^{(T)} (Q)} 
\right] . 
\label{G2} 
\end{eqnarray} 
The replacement (\ref{kiso}) and Fig.~3 tell us \cite{nie} that 
Eq.~(\ref{sof}) describes the differential rate for a set of 
processes, in which real and/or virtual photon(s) participate. It is 
to be noted that the (real) photons in the QED plasma are in thermal 
equilibrium. Then, the photon(s) does not come out of the plasma, so 
that, when the decay neutrino goes out from the plasma, it does not 
accompany photon(s). [In this relation, see \cite{olivo}]. 

In a similar manner, we obtain, for the contribution from the 
processes (\ref{pro1}) and (\ref{pro2}), 
\begin{eqnarray} 
\frac{d \tilde{\Gamma}^{(2)}_p}{d q_0} & = & - \frac{G^2}{\pi^2} 
\frac{1}{e^2} \frac{1}{k^2} \left[ \theta (- q_0) + n_B (|q_0|) 
\right] \epsilon (q_0 - k) \nonumber \\ 
&& \times \int_{\cal R} d q \, q \, \bar{f}_{\nu'} \left( X; k - 
q_0, \left( 1 - \frac{q_0}{k} + \frac{Q^2}{2 k^2} \right) {\bf k} 
\right) G^{(2)} (Q) . 
\label{sof1} 
\end{eqnarray} 
\narrowtext 

\noindent 
The diagram for this is the same as Fig.~3, provided that the type-1 
vertex and the type-2 vertex in Fig.~3 are interchanged. 

In the next section, we shall use the formulae displayed above for 
the whole $Q^2$-region. 
\subsection{Net decay rate} 
The net decay rate $\Gamma_d^{\mbox{\scriptsize{net}}}$ and the net 
production rate $\Gamma_p^{\mbox{\scriptsize{net}}}$ are 
\begin{eqnarray} 
\Gamma_d^{\mbox{\scriptsize{net}}} (X; {\bf k}) & = & N_{\nu} (X; 
{\bf k}) \Gamma_d (X;{\bf k}) \nonumber \\ 
&& - [ 1 - N_{\nu} (X; {\bf k})] \Gamma_p 
(X; {\bf k}) \nonumber \\ 
& = & - \Gamma_p^{\mbox{\scriptsize{net}}} (X; {\bf k}) , 
\label{nett} 
\end{eqnarray} 
where $N_{\nu} (X; {\bf k})$ is an (on mass-shell) distribution 
function of $\nu$. When $\nu$, $\nu'$, and $\bar{\nu}'$ are in 
thermal and chemical equilibrium, $N_\nu$, $N_{\nu'}$, and 
$N_{\bar{\nu}'}$ take similar form to Eq.~(\ref{bunp}). Using 
Eq.~(\ref{kan}), one can show, in this case, that the detailed 
balance holds, $\Gamma_d^{\mbox{\scriptsize{net}}} (X; {\bf k}) 
\to \Gamma_d^{\mbox{\scriptsize{net}}} (k) = 0$. 
\subsection{Procedure of determining (anti)neutrino distribution 
functions} 
Here we study the neutrino or weak-interaction sector. As discussed 
above after Eq.~(\ref{Adesu}), the nonleading piece of 
$\hat{S}_{\nu'}$, $\hat{S}_{\nu'}^{(1)}$, is $O (G^2)$ smaller than 
the leading piece $\hat{S}_{\nu'}^{(0)}$ and may be neglected. 
Higher-order corrections to $\Sigma^{(\nu)}_{2 1}$, 
Eq.~(\ref{sora}), and to $\Sigma^{(\nu)}_{1 2}$, Eq.~(\ref{mori}), 
due to weak interaction, are also $O (G^2)$ smaller than the 
respective leading contributions analyzed above, so that they can be 
ignored. There is, however, an exception to this argument. Our 
formalism outlined in Appendix A accompanies, for each neutrino 
$\nu$, the evolution equation for $f_\nu$, which is included in the 
neutrino propagator $\hat{S}_\nu (X; P)$: 
\begin{eqnarray}
{\partial\mbox{\hspace*{0.3mm}}\kern-0.1em\raise0.3ex\llap{/}
\kern0.15em\relax}_X f_\nu (X; P) L & = & F_\nu (X; P) , 
\label{kj} \\ 
F_\nu (X; P) & = & i \left[ (1 - f_\nu) \Sigma_{1 2}^{(\nu)} + f_\nu 
\Sigma_{2 1}^{(\nu)}\right] . 
\end{eqnarray}
Here $\Sigma_{i j}^{(\nu)}$ is the $(i j)$ element of the 
self-energy-part matrix $\hat{\Sigma}_\nu$ and is of $O (G^2)$. For 
computing $F_\nu (X; P)$, all the relevant contributions should be 
included, among which are the contributions analyzed above. (See 
Appendix A for more details.) 

We show in Appendix A that, on the mass-shell $p_0 = \pm p$, 
Eq.~(\ref{kj}) turns out to the Boltzmann equation (\ref{yama}) and 
its relatives, Eqs.~(\ref{kure1}) and (\ref{sato}). As a matter of 
fact, as far as $O (G^2)$ contributions are concerned, only $f_\nu$ 
on the mass-shell is relevant because of the presence of $\delta 
(P^2)$ in Eq.~(\ref{zero}). Eq.~(\ref{kj}) describes the spacetime 
evolution of $f_\nu (X; P)$ under a given initial data $f_\nu 
(X_{\mbox{\scriptsize{in}}}^0, {\bf X}; P )$ at an initial time $X^0 
= X_{\mbox{\scriptsize{in}}}^0$. Although the RHS of Eq.~(\ref{kj}) 
is of $O (G^2)$, its effect cannot be ignored in general. This is 
because, in the process of solving Eq.~(\ref{kj}) for $f_\nu (X; 
P)$, integration over {\em large spacetime scale} is involved, so 
that the $O (G^2)$ effect is enhanced. 

In conclusion, neutrino self-energy part can be ignored everywhere 
but in $F_\nu$ in Eq.~(\ref{kj}). In computing the reaction rates, 
as dealt with here, one should substitute the (anti)neutrino 
distribution functions, which are determined through 
Eq.~(\ref{kj}) in a self-consistent manner. This applies to every 
neutrino. 
\section{Numerical computation} 
In this section we present a result of numerical computation of 
the \lq\lq portion'' of $d \tilde{\Gamma}_{d (p)} / d q_0$ and 
$\tilde{\Gamma}_{d (p)}$ that are independent of the $\nu'$ 
distribution function. We are interested in the type-II supernova 
environment, which is a QED plasma whose core temperature is $T 
\sim 30 - 60$ MeV and electron chemical potential is $\mu \sim 
350$ MeV \cite{le-b}. 
\subsection{Differential reaction rates} 
In general, the number-distribution function of $\nu'$ 
($\bar{\nu}'$), 
$N_{\nu' (\bar{\nu}')} (X; k_0', {\bf k}')$ $(K' = K - Q)$, is 
anisotropic, and one should compute Eqs.~(\ref{1221}), 
(\ref{prod}), (\ref{sof}), and (\ref{sof1}) substituting $N_{\nu'}$ 
and $N_{\bar{\nu}'}$, which are to be determined self consistently. 
In this section, we restrict ourselves to the case of isotropic 
distribution, $N_{\nu' (\bar{\nu}')} (k_0', {\bf k}') = N_{\nu' 
(\bar{\nu}')} (k_0')$. In this case, Eqs.~(\ref{1221}), 
(\ref{prod}), (\ref{sof}), and (\ref{sof1}) may be written in the 
form, 
\begin{eqnarray*} 
\frac{d \tilde{\Gamma}_d^{(i)}}{d q_0} & = & \frac{G^2}{\pi^2} 
\frac{1}{e^2} \frac{1}{k^2} \left[ 1 - f_{\nu'} (X; k - q_0) \right] 
\nonumber \\ 
&& \times {\cal G}^{(i)} (q_0, k, T) \mbox{\hspace*{21ex}} (i = 1, 
2) , 
\nonumber \\ 
\frac{d \tilde{\Gamma}_p^{(i)}}{d q_0} & = & \frac{G^2}{\pi^2} 
\frac{1}{e^2} \frac{e^{- q_0 / T}}{k^2} f_{\nu'} (X; k - q_0) {\cal 
G}^{(i)} (q_0, k, T) \nonumber \\ 
&& \mbox{\hspace*{36ex}} (i = 1, 2) . 
\end{eqnarray*} 
Here 
\begin{eqnarray*} 
{\cal G}^{(i)} (q_0, k, T) & \equiv & \epsilon (k - q_0) [\theta 
(q_0) + n_B (|q_0|)] \nonumber \\ 
&& \times \int_{\cal R} d q \, q \, G^{(i)} (Q) \;\;\;\;\;\;\;\;\;
\;\;\;\;\;\; (i = 1, 2) 
\end{eqnarray*} 
with $G^{(1)} (Q)$ and $G^{(2)} (Q)$ as in Eqs.~(\ref{G1}) and 
(\ref{G2}), respectively. We compute ${\cal G}^{(i)} (q_0, k, T)$ 
($i = 1, 2$) for various values for the parameters $T$ and $k$. For 
the chemical potential, unless otherwise stated, we take $\mu = 350$ 
MeV \cite{le-b}. 

In Figs.~4~-~8, we display the results of numerical computation for 
different values for $k$ and $T$. The solid lines represent the 
total contributions ${\cal G} \equiv {\cal G}^{(1)} + {\cal 
G}^{(2)}$, while the dot-dashed lines represent ${\cal G}^{(1)}$. 
The figures \lq\lq (a)'' display ${\cal G}$ and ${\cal G}^{(1)}$ in 
the region $q_0 \leq k$ [the region of the processes 
(\ref{process1}) and (\ref{pro1})] and the figures \lq\lq (b)'' 
display ${\cal G}$ and ${\cal G}^{(1)}$ in the region $q_0 > k$ [the 
region of the processes (\ref{process2}) and (\ref{pro2})]. Some 
observations are in order. 
\begin{itemize} 
\item Figures 4~-~7 show the results for different values of $k$ 
with $T = 50$ MeV. We see, as is expected, that for smaller 
incident-neutrino energy $k$, ${\cal G}^{(2)} / {\cal G}$ is larger. 
In the region of figures (b) [the region of the processes 
(\ref{process2}) and (\ref{pro2})], both ${\cal G}^{(1)}$ and ${\cal 
G}^{(2)}$ ($= {\cal G} - {\cal G}^{(1)}$) are positive. In the 
region of figures (a) [the region of the processes (\ref{process1}) 
and (\ref{pro1})], except for the small region $q_0 \sim 0$ (or $k' 
\sim k$) in the case of relatively small $k / T$, ${\cal G}^{(2)}$ 
is negative. Referring to the reaction-rate formula \cite{nie}, one 
can see what kind of physical processes are involved in $d 
\tilde{\Gamma}_d^{(2)} / d q_0$ (Eq.~(\ref{sof})) and $d 
\tilde{\Gamma}_p^{(2)} / d q_0$ (Eq.~(\ref{sof1})). As a matter of 
fact, each of them involves a set of infinite number of coherent 
processes. A few examples of them that are involved in $d 
\tilde{\Gamma}_d^{(2)} / d q_0$ are 
\[ 
\nu + e \to e + \gamma + \nu' \, , \;\;\;\;\; \nu + e \to e + e + 
e^+ + \nu' . 
\] 
Figures~4~-~8 tell us that, for most regions displayed in figures 
(a), an infinite number of \lq\lq interference contributions'' is 
summed up to be negative, so that $d \tilde{\Gamma}_d^{(2)} / d q_0$ 
and $d \tilde{\Gamma}_p^{(2)} / d q_0$ are negative. 
\item 
Both in figures (a) and (b), ${\cal G}^{(2)} (q_0, k, T)$ exhibits 
peak structure. For figures (a) [$q_0 < k$], the peak is at $q_0 
\simeq 0$ or $k' = k - q_0 \simeq k$ and is more prominent for 
smaller incident energy $k$. The structure of figures (b) [$q_0 \geq 
k$] may be understood as follows. In the hard-thermal-loop 
approximation [cf. Eq.~(\ref{kinnji})], 
$\displaystyle{\raisebox{0.6ex}{\scriptsize{*}}} \! \Delta^{(T / 
L)}_F (Q) \simeq 1 / [Q^2 - F_{T / L} (Q)]$ (see Eqs.~(\ref{effe1}) 
and (\ref{dai})). Then, as is well known \cite{le-b} or as can be 
shown from Eq.~(\ref{effe1}) with Eqs.~(\ref{FT}) and (\ref{FL}), 
$\displaystyle{\raisebox{0.6ex}{\scriptsize{*}}} \! \Delta_{T / 
L}^{(1 2) / (2 1)} (Q)$ in Eq.~(\ref{a11}) turns out to be  of the 
form 
\begin{eqnarray} 
\displaystyle{\raisebox{0.6ex}{\scriptsize{*}}} \! \Delta_{T / 
L}^{(1 2) / (2 1)} (Q) & = & 2 i [\theta (\mp q_0) + n_B (|q_0|)] 
Z_{T / L} (q) \nonumber \\ 
&& \times \delta (q_0 - \omega_{T / L} (q)) \mbox{\hspace*{4ex}} 
(q_0 > q). 
\nonumber \\ 
\label{obss} 
\end{eqnarray}
The dispersion curves, $q_0 = \omega_T (q)$ and  $q_0 = \omega_L 
(q)$, are schematically shown in Fig.~2. Use of the actual 
$\displaystyle{\raisebox{0.6ex}{\scriptsize{*}}} \! \Delta^{T / L}_F 
(Q) = 1 / [Q^2 - \Pi_F^{T / L} (Q)]$ results in the change of 
$\delta (q_0 - \omega_{T / L} (q))$ in Eq.~(\ref{obss}) to the 
functions with finite width that are (sharply) peaked at $q_0 \simeq 
\omega_{T / L} (q)$. Inspection of Fig.~2 with these observation in 
mind allows us to understand the structure of figures (b). 
\end{itemize} 

For the purpose of seeing the effect of the chemical potential 
$\mu$, we display in Fig.~9 the result for $(k, T, \mu) = (10, 50, 
0)$ MeV. [For the QED plasma in the early universe, $\mu \simeq 0$.] 
We see that ${\cal G}^{(2)} << {\cal G}^{(1)}$, so that the peak 
structure is less prominent when compared to the case of $\mu = 350$ 
MeV, Fig.~5. ${\cal G}$ in the region $q_0 > k$ is much larger than 
${\cal G}$ in the region $q_0 < k$.  

Above computation is carried out neglecting the electron mass $m_e$. 
Inclusion of the electron mass $m_e$ causes a change in $\Pi_F^{(S)} 
(Q)$ $(S = T, L, V\!A )$ in the region $|Q^2| \leq O (m_e^2)$. For 
the purpose of getting a measure to what extent the approximation 
$m_e = 0$ is good one, we perform all numerical computations by 
simply cutting off the region $|Q^2| < m_e^2$. This cutoff turns 
out to reduce ${\cal G}^{(i)}$ ($i = 1, 2$). Dashed lines in 
Figs.~4~-~9 show the result of computation. In most regions of 
Figs.~4~-~9, no substantial reduction arises. Especially, for the 
region $q_0 \geq k$, no sizable reduction arises and we do not 
display the results in figures (b). For the region $q_0 < k$, 
prominent reduction occurs only at $q_0 \simeq 0$, at which ${\cal 
G}^{(2)}$ peaks. Larger reduction occurs for smaller $k$. 
\subsection{Decay rate} 
For computing the contributions to the decay or damping rate 
$\tilde{\Gamma}_d$ $(= \tilde{\Gamma}_d^{(1)} + 
\tilde{\Gamma}_d^{(2)})$ (cf. Eq.~(\ref{gam})) and to the production 
rate $\tilde{\Gamma}_p$ $(= \tilde{\Gamma}_p^{(1)} + 
\tilde{\Gamma}_p^{(2)})$ (see Eq.~(\ref{rate1})), knowledge for the 
distribution functions, $N_{\nu'}$ and $N_{\bar{\nu}'}$, is 
necessary. Furthermore, for computing the net decay rate 
$\Gamma_d^{\mbox{\scriptsize{net}}} (X; {\bf k})$, Eq.~(\ref{nett}), 
knowledge for the distribution function $N_\nu (X; {\bf k})$ is 
necessary. 

Here we compute the damping rate $\tilde{\Gamma}_d$ of an incident 
$\nu$ on a hot and dense QED plasma with no background neutrinos, 
$N_\nu = N_{\nu'} = N_{\bar{\nu}} = N_{\bar{\nu}'} = 0$. Then, the 
process (\ref{process2}) is absent. Displayed in Figs.~10 and 11 are 
the total contribution $\tilde{\Gamma}_d$ $(= \tilde{\Gamma}^{(1)}_d 
+ \tilde{\Gamma}^{(2)}_d)$ and the partial contribution 
$\tilde{\Gamma}^{(1)}_d$ for $\mu = 350$ MeV and, in respective 
order, $T = 50$ MeV and $20$ MeV. Figures 12 and 13 show the result 
for $\mu = 0$ and, in respective order, $T = 50$ MeV and $20$ MeV. 

Cutting off the contribution from the region $|Q^2| < m_e^2$ does 
not result in sizable reduction. 

Characteristic features: 
\begin{itemize} 
\item For the range of $k$, $T$, and $\mu$ displayed in 
Figs.~10~-~13, the contribution from the soft-$Q$ region, 
$\tilde{\Gamma}^{(2)}_d$, is not very large. 
\item Figures 12 and 13 tell us that, for $\mu = 0$, 
$\tilde{\Gamma}_d$ is almost linear in $k$. As a matter of fact, 
$\tilde{\Gamma}_d$'s in Figs.~12 and 13 are well parametrized as 
\[ 
\tilde{\Gamma}_d (k, T) = c G^2 k^{1 + \alpha} T^{4 - \alpha} 
\] 
with $(c, \alpha) = (0.60, 0.04)$. 
\end{itemize} 
\section{Discussions} 
On the basis of the formalism, outlined in Appendix A, for dealing 
with nonequilibrium quantum-field systems, we have derived the 
reaction-rate formulae for neutrino-conversion process and its 
relatives, the processes which occur in the medium that consists of 
a hot and dense QED plasma and background neutrinos. The formalism 
involves the Boltzmann equation and two related ones, which describe 
spacetime evolution of (anti)neutrino distribution function. 
Illustrative computations of the differential reaction rate $d 
\Gamma_{d (p)} / d q_0$ is made in Sec.~IIIA and of the total decay 
rate  $\Gamma_d$ in Sec.~IIIB. 

For relativistic particles dealt with here, the mean-free path $l$ 
is related to the decay rate $\Gamma$ through $l = 1 / \Gamma$ 
\cite{disp}, which, in turn, is related to the imaginary part of the 
refractive index $\mbox{Im} [n] = (2 l k)^{- 1} = \Gamma / 2 k$. 
Computation in Sec.IIIB shows that, in the range of Figs.~10 and 11, 
order of magnitude of $\tilde{\Gamma}_d$ is $10^{-15} \sim 10^{- 
12}$ MeV. Then, we see from Eq.~(\ref{rate00}) that $l \simeq [0.2 
\sim 200] / |U_{e \nu}^* U_{e \nu'}|^2$ m, which is much less than 
the core size of the type-II supernova. This means that, when 
applying to the actual supernova, $\bar{\nu} \nu'$ as well as $\nu 
\bar{\nu}'$ production processes are also important and, through 
these processes, (anti)neutrinos are produced. Thus, one has to 
perform an analysis by the full use of the formalism in Appendix A. 
The evolution of the neutrino- and antineutrino-distribution 
functions should be dealt with through the Boltzmann equation and 
its relatives. Once the initial distribution function of 
(anti)neutrinos are given, the framework presented here allows one 
to determine hereafter of the system. Concrete numerical analysis 
along this line is outside of the scope of the present paper. 
\section*{Acknowledgments}
The authors thank T. Hatsuda for useful comments. M. K. contributed 
to this work at the early stage. A. Ni\'egawa was supported in part 
by a Grant-in-Aide for Scientific Research ((C)(2) (No.~12640287)) 
of the Ministry of Education, Science, Sports and Culture of Japan. 
\begin{appendix} 
\setcounter{section}{0}
\section{Nonequilibrium neutrino propagator and \lq\lq healthy'' 
perturbation theory} 
\def\theequation{\mbox{\Alph{section}\arabic{equation}}} 
A perturbative framework for dealing with massless Dirac fermions in 
a nonequilibrium quantum-field system is presented in \cite{niefer}. 
Extension of the framework to the case of massless left-handed 
neutrino $\nu_L$ is straightforward. Here we briefly describe 
somewhat simpler framework for $\nu_L$, which is sufficient for the 
present purpose. We employ the derivative expansion and use the 
gradient approximation throughout. 
\subsection{Free propagator} 
We start with introducing a standard form \cite{ume,niefer} for 
nonequilibrim matrix propagator of a massless left-handed neutrino 
$\nu_L$: 
\begin{eqnarray}
\hat{S}_\nu (x, y) & = & \int \frac{d^{\, 4} u}{(2 \pi)^4} \int
\frac{d^{\, 4} v}{(2 \pi)^4} \hat{B}_L (x, u) \nonumber \\ 
&& \mbox{\hspace*{11ex}} \times \hat{S}_{RA} (u - v) 
\hat{B}_R (v, y) , 
\label{dea} \\ 
\hat{B}_L (x, y) & = & \left( 
\begin{array}{cc} 
\delta^{\, 4} (x - y)  & \; - f_\nu (x, y) \\ 
\delta^{\, 4} (x - y) & \; \delta^{\, 4} (x - y) - f_\nu (x, y) 
\end{array} 
\right) , \nonumber \\ 
\hat{B}_R (x, y) & = & \left( 
\begin{array}{cc} 
\delta^{\, 4} (x - y) - f_\nu (x, y) & \; - f_\nu (x, y) \\ 
\delta^{\, 4} (x - y) & \; \delta^{\, 4} (x - y) 
\end{array} 
\right) , \nonumber \\ 
\hat{S}_{RA} & = & \mbox{diag} \left( S_R, - S_A \right) . \nonumber 
\end{eqnarray}
Here \lq\lq $\,\, \hat{} \, \,$'' denotes a $2 \times 2$ matrix in 
a \lq\lq type space.'' In a $4 \times 4$ Dirac-matrix space, 
$\hat{B}_L$ and $\hat{B}_R$ are the unit matrices. The Fourier 
transform of $S_{R (A)}$ reads 
\begin{eqnarray}
S_{R (A)} (P) & = & L 
P\kern-0.1em\raise0.3ex\llap{/}\kern0.15em\relax \Delta_{R (A)} 
(P) \;\;\;\;\; (L = (1 - \gamma_5) / 2), \nonumber 
\end{eqnarray}
where $\Delta_{R (A)} (P)$ is as in Eq.~(\ref{ra}) and $f_\nu (x, 
y)$ is the inverse Wigner transform of 
\begin{eqnarray*} 
f_\nu (X; P) & = & \theta (p_0) N_\nu (X; p_0, {\bf p}) \\ 
&& + \theta (- p_0) \left[ 1 - N_{\bar{\nu}} (X; |p_0|, - {\bf p}) 
\right] 
\end{eqnarray*} 
with $X = (x + y) / 2$. Here $N_\nu$ $(N_{\bar{\nu}})$ is the 
number-density function of $\nu$ ($\bar{\nu}$). Computation of 
Eq.~(\ref{dea}) to the gradient approximation yields 
\begin{eqnarray}
\hat{S}_\nu (x, y) & = & \int \frac{d^{\, 4} P}{(2 \pi)^4} e^{- i P 
\cdot (x - y)} \hat{S}_\nu \left( \frac{x + y}{2}; P \right) , 
\nonumber \\ 
\hat{S}_\nu (X; P) & = & \hat{S}^{(0)}_\nu (X; P)  + 
\hat{S}^{(1)}_\nu (X; P) . \nonumber 
\end{eqnarray} 
The leading term $\hat{S}^{(0)}_\nu$ is given by Eq.~(\ref{zero}) 
with $\nu$ for $\nu'$ and the nonleading term $\hat{S}_\nu^{(1)}$ 
reads 
\begin{equation}
\hat{S}^{(1)}_\nu = - i L \hat{A}_+ \left[ \left( 
{\partial\mbox{\hspace*{0.3mm}}\kern-0.1em\raise0.3ex\llap{/}
\kern0.15em\relax}_X f_\nu \right) \frac{\bf P}{P^2} + 2 \left( P 
\cdot \partial_X f_\nu \right) 
P\kern-0.1em\raise0.3ex\llap{/}\kern0.15em\relax 
\frac{\partial}{\partial P^2} \frac{\bf P}{P^2} \right] , 
\label{nupro2} 
\end{equation} 
where $\hat{A}_+$ is as in Eq.~(\ref{Adesu}) and ${\bf P}$ denotes 
principal part. 
\subsection{Free and counter actions} 
The propagator introduced above is an inverse of the kernel of a 
free action $A_0$, which we now find. Applying 
${\partial\mbox{\hspace*{0.3mm}}\kern-0.1em\raise0.3ex\llap{/}
\kern0.15em\relax}_x$ to $\hat{S}_\nu (x, y)$ in Eq.~(\ref{dea}), 
we find the form of $A_0$, to the gradient approximation: 
\begin{eqnarray}
A_0 & = & \int d^{\, 4} x \, \bar{\Psi} (x) i 
{\partial\mbox{\hspace*{0.3mm}}\kern-0.1em\raise0.3ex\llap{/}
\kern0.15em\relax} L \Psi (y) - A_C , 
\label{A0} 
\\ 
\bar{\Psi} & = & \left( \bar{\psi}_1, \bar{\psi}_2 \right) , 
\mbox{\hspace*{10ex}} \Psi (x) = \left( 
\begin{array}{c} 
\psi_1 \\ 
\psi_2 
\end{array} 
\right) , 
\label{tate} \\ 
A_c & = & \int \frac{d^{\, 4} x}{(2 \pi)^4} \int 
\frac{d^{\, 4} y}{(2 \pi)^4} \bar{\Psi} (x) \nonumber \\ 
&& \mbox{\hspace*{8ex}} \times 
(i {\partial\mbox{\hspace*{0.3mm}}\kern-0.1em\raise0.3ex\llap{/}
\kern0.15em\relax}_x + 
i {\partial\mbox{\hspace*{0.3mm}}\kern-0.1em\raise0.3ex\llap{/}
\kern0.15em\relax}_y ) L f_\nu (x, y) \hat{A}_- \Psi (y) , 
\end{eqnarray}
where $\hat{A}_-$ is as in Eq.~(\ref{Adesu}). In Eq.~(\ref{tate}), 
the subscripts of the field denote the type of field in real-time 
formalism \cite{chou}. From Eq.~(\ref{A0}), we see that the action 
of the theory turns out to be 
\[ 
A = A_0 + A' + A_c , 
\] 
where $A'$ includes other fields than $\nu_L$ and $\bar{\nu}_L$ and 
the interactions between fields. It is to be noted that the counter 
action $A_c$ appears. 
\subsection{Self-energy part and \lq\lq healthy'' perturbation 
theory} 
The self-energy-part matrix of $\nu$ reads 
\[ 
\hat{\Sigma}_\nu (X; P) = 
\hat{\Sigma}_\nu^{(\mbox{\scriptsize{loop}})} (X; P) - i 
{\partial\mbox{\hspace*{0.3mm}}\kern-0.1em\raise0.3ex\llap{/}
\kern0.15em\relax}_X f (X; P) L \hat{A}_- , 
\] 
where $\hat{\Sigma}_\nu^{(\mbox{\scriptsize{loop}})}$ comes from 
loop diagrams and the second term on the RHS comes from the counter 
action $A_c$. Computation of a $\hat{\Sigma}$-inserted propagator 
to the leading order of derivative expansion yields 
\begin{eqnarray}
&& \hat{S}_\nu (X; P) \hat{\Sigma}_\nu (X; P) \hat{S}_\nu (X; P) 
\nonumber \\ 
&& \mbox{\hspace*{5ex}} = - i S_R \left[ 
\left( {\partial\mbox{\hspace*{0.3mm}}\kern-0.1em\raise0.3ex\llap{/}
\kern0.15em\relax}_X f_\nu \right) 
L \right. \nonumber \\ 
&& \mbox{\hspace*{5ex}} \left. - i \left\{ (1 - f_\nu) \Sigma_{1 
2}^{(\nu)} + f_\nu \Sigma_{2 
1}^{(\nu)} \right\} \right] S_A \, \hat{A}_+ \nonumber \\ 
&& \mbox{\hspace*{8ex}} + ... . \nonumber 
\end{eqnarray}
Here $f_\nu = f_\nu (X; P)$ and \lq\lq $...$'' contains the terms 
with $S_R \Sigma_R^{(\nu)} S_R$ and with $S_A \Sigma_A^{(\nu)} S_A$, 
where $\Sigma^{\nu}_{R (A)} = \Sigma^{\nu}_{1 1} + \Sigma^{\nu}_{12 
(21)}$ is the retarded (advanced) self-energy part \cite{chou}. 
Observing $S_R S_A \propto 1 / \left[ (P^2 + i 0^+) (P^2 - i 0^+) 
\right]$, we see that $S_R S_A$ possesses pinch singularities at 
$p_0 = \pm p$ in a complex $p_0$-plane. Then, by 
demanding\footnote{As a matter of fact, demanding Eq.~(\ref{kk}) to 
hold in any region(s) of $P$ will do, as far as $p_0 = + p$ and $p_0 
= - p$ are within that region(s).} 
\begin{eqnarray}
{\partial\mbox{\hspace*{0.3mm}}\kern-0.1em\raise0.3ex\llap{/}
\kern0.15em\relax}_X f_\nu (X; P) L & = & F_\nu (X; P) , \nonumber 
\\ 
F_\nu (X; P) & = & i \left[ (1 - f_\nu) \Sigma_{1 2}^{(\nu)} + f_\nu 
\Sigma_{2 1}^{(\nu)}\right] , 
\label{kk} 
\end{eqnarray}
we attain a pinch-singularity free perturbation theory. 
\subsection{Boltzmann equation and its relative equations} 
After multiplying $L 
P\kern-0.1em\raise0.3ex\llap{/}\kern0.15em\relax$ from the left of 
Eq.~(\ref{kk}), we take a trace and go on the mass-shell $p_0 = \pm 
p$. Referring to Eqs.~(\ref{sora}) and (\ref{mori}) together with 
their antiparticle counterparts, we obtain, with obvious notation, 
\begin{eqnarray}
&& (\partial_{X_0} + {\bf v} \cdot \nabla_X) N_\pm (X; p, {\bf 
p}) \nonumber \\ 
&& \mbox{\hspace*{5ex}} = \left( 1 - N_{\pm} (X; p, {\bf p}) 
\right) \Gamma_p^{(\pm)} (X; {\bf p}) \nonumber \\ 
&& \mbox{\hspace*{7.5ex}} -  N_{\pm} (X; p, {\bf p}) 
\Gamma_d^{(\pm)} (X; {\bf p}) . 
\label{yama} 
\end{eqnarray}
Here ${\bf v} \equiv {\bf p} / p$ and \lq\lq $+$'' [\lq\lq $-$''] 
stands for $\nu$ [$\bar{\nu}$]. As can be seen from 
Eq.~(\ref{nett}), the RHS is the net production rate of 
$\nu/\bar{\nu}$ and Eq.~(\ref{yama}) is nothing but the Boltzmann 
equation. 

Similarly, multiplications of $L \gamma^0$ and and of $L 
\vec{\gamma}_\perp$ ($= L \vec{\gamma} - L (\vec{\gamma} \cdot 
\hat{\bf p}) \hat{\bf p}$) to Eq.~(\ref{kk}) yields, in respective 
order, 
\begin{eqnarray}
&& \partial_{X_0} N_\pm (X; p, {\bf p}) = \pm \frac{1}{2} \, 
\mbox{Tr} \left[ L \gamma^0 F_\nu \right]_{p_0 = \pm p, {\bf p} \to 
\pm {\bf p}} , 
\label{kure1} 
\\ 
&& \vec{\gamma}_\perp \cdot \nabla_X N_\pm (X; p, {\bf p}) = 
\pm \frac{1}{2} \, \mbox{Tr} \left[ L \vec{\gamma}_\perp F_\nu 
\right]_{p_0 = \pm p, {\bf p} \to \pm {\bf p}} . \nonumber \\ 
\label{sato} 
\end{eqnarray}
If (initial) distributions are spatially isotropic, the RHS of 
Eq.~(\ref{sato}) vanishes. 
\section{Self-energy-part resummed propagator of a photon} 
Elements of the self-energy-part-resummed photon 
prop\-a\-ga\-tor matrix (in Landau gauge) reads \cite{le-b,pis,lan} 
\begin{eqnarray} 
\displaystyle{\raisebox{0.6ex}{\scriptsize{*}}} \! \Delta^{\alpha 
\beta}_{i j} (Q) & = & - {\cal P}_T^{\alpha \beta} (\hat{\bf q}) \, 
\displaystyle{\raisebox{0.6ex}{\scriptsize{*}}} \! \Delta^{(i j)}_T 
(Q) - {\cal P}_L^{\alpha \beta} (Q) \, 
\displaystyle{\raisebox{0.6ex}{\scriptsize{*}}} \! \Delta^{(i j)}_L 
(Q) \nonumber \\ 
&& \mbox{\hspace*{24ex}} (i, j = 1, 2) , 
\label{effe} 
\end{eqnarray} 
where ${\cal P}_{T}^{\alpha \beta} (\hat{\bf q})$ and ${\cal 
P}_{L}^{\alpha \beta} (Q)$ are as in Eq.~(\ref{proj}) and 
\begin{eqnarray}
\displaystyle{\raisebox{0.6ex}{\scriptsize{*}}} \! \Delta^{(1 
1)}_{T / L} (Q) & = & - \left( \displaystyle{ 
\raisebox{0.6ex}{\scriptsize{*}}} \! \Delta^{(2 2)}_{T / L} (Q) 
\right)^* \nonumber \\ 
& = & \displaystyle{ \raisebox{0.6ex}{\scriptsize{*}}} \! 
\Delta_F^{(T / L)} (Q) + 2 i n_B (|q_0|) \, \mbox{Im} \displaystyle{ 
\raisebox{0.6ex}{\scriptsize{*}}} \! \Delta_F^{(T / L)} (Q) , 
\nonumber \\ 
\displaystyle{\raisebox{0.6ex}{\scriptsize{*}}} \! \Delta^{(1 2) / 
(2 1)}_{T / L} (Q) & = & 2 i [\theta (\mp q_0) + n_B (|q_0|)] \, 
\mbox{Im} \displaystyle{ \raisebox{0.6ex}{\scriptsize{*}}} \! 
\Delta_F^{(T / L)} (Q) , \nonumber \\ 
\label{a11} 
\\ 
\displaystyle{\raisebox{0.6ex}{\scriptsize{*}}} \! \Delta_F^{(T /L)} 
(Q) & = & \frac{1}{Q^2 - \Pi_F^{(T / L)} (Q)} . 
\label{effe1} 
\end{eqnarray} 
$\Pi^{(T / L)}_F$ is computed to the one-loop order in Appendix C. 
\setcounter{section}{2} 
\section{Self-energy part $\Pi_F^{(S)}$ $(S= T, L, 
V\!A)$} 
\def\theequation{\mbox{\Alph{section}\arabic{equation}}} 
Here we compute the lowest-order contribution to $\Pi_F^S (Q)$ $(S = 
T, L, V\!A)$), Eq.~(\ref{PiF}). We are interested in the high-$T$ 
and large-$\mu$ region $T, \mu >> m_e$, and then we ignore $m_e$. 
The effect of $m_e$ $(\neq 0)$ is discussed in Sec. III. 
\subsection*{Computation of $\Pi_F^{(S)} (Q)$ $(S = T, L, V\!A$)} 
We decompose $\Pi^{(T)}_F (Q)$ and $\Pi^{(L)}_F (Q)$ into three 
parts, 
\begin{equation} 
\Pi_F^{(T / L)} (Q) = F_{T / L}^{(0)} (Q) + F_{T / L} (Q) + G_{T / 
L} (Q) . 
\label{dai} 
\end{equation} 
$F_{T / L}^{(0)}$ stands for the vacuum contribution and $F_{T / L}$ 
stand for the contributions that dominate in the soft-$Q$ region, 
the latter contributions which are called hard thermal loop 
\cite{le-b,pis}. Incidentally, $\Pi_F^{(V\!A)} (Q)$ has no hard 
thermal loop. 

Straightforward computation of Eq.~(\ref{PiF}) (cf. 
Eqs.~(\ref{W-pol1}), (\ref{decomp2}), and (\ref{proj})) using 
Eqs.~(\ref{esu}) and (\ref{bunp}) yields 
\begin{eqnarray}
F_T^{(0)} (Q) & = & F_L^{(0)} (Q) = - \frac{\alpha}{3 \pi} Q^2 
\left[ \frac{5}{3} - \ln \left(\frac{- Q^2}{\mu_r^2} \right) 
\right] , 
\label{vacc} 
\\ 
F_T (Q) & = & \frac{3}{2} \, m_\gamma^2 \frac{q_0}{q} 
\left[ \frac{q_0}{q} - \frac{Q^2}{2 q^2} \, \ln 
\frac{q_0 + q}{q_0 - q} \right] , 
\label{FT} \\ 
F_L (Q) & = & - 3 \, m_\gamma^2 \frac{Q^2}{q^2} 
\left[ 1 - \frac{q_0}{2 q} \, \ln \frac{q_0 + q}{q_0 - q} \right] 
, 
\label{FL} \\ 
G_T (Q) & = & - \frac{\alpha}{\pi} \, Q^2 \left[ \left( 1 + 
\frac{Q^2}{2 q^2} \right) I_1 + \frac{2}{q^2} \left( I_2  - q_0 
I_3 \right) \right] , \nonumber \\ 
\\ 
G_L (Q) & = & \frac{\alpha}{\pi} \, \frac{Q^2}{q^2} \left[ Q^2 I_1 + 
4 \left( I_2 - q_0 I_3 \right) \right] , 
\end{eqnarray} 
and 
\begin{equation} 
\Pi^{(V \! A)}_F = - \frac{e^2}{4 \pi^2} \frac{Q^2}{q^3} 
\left[ \mu q_0 \ln \frac{q_0 + q}{q_0 - q} - q q_0 \tilde{I}_3 + 2 
q \tilde{I}_1 \right] . 
\label{13} 
\end{equation} 
In obtaining the vacuum contribution (\ref{vacc}), we have used the 
$\overline{\mbox{MS}}$ scheme\footnote{We have adopted a convention 
that Dirac gamma matrices are $4 \times 4$ matrices in 
$D$-dimensional spacetime.} and $\mu_r$ is the renormalization 
scale, for which we choose $\mu_r = \sqrt{T^2 + \mu^2}$. 
Incidentally, the vacuum part of $\Pi^{(V\! A)}_F$ vanishes. In the 
above equations, $\alpha = e^2 / 4 \pi$, $m_\gamma^2 = e^2 (T^2 + 3 
\mu^2 / \pi^2) / 9$ is the thermal mass of an electron, and 
\begin{eqnarray}
I_1 & \equiv & - \frac{1}{2 q} \int_0^\infty d p [n_+ (p) + n_- 
(p)] \, \ln 
\left( \frac{L_{+ +} \, L_{- -}}{L_{+ -} \, L_{- +}} \right) , 
\nonumber \\ 
&& \\ 
I_2 & \equiv & \frac{\pi^2}{6} \, \left(T^2 + \frac{3 \mu^2}{\pi^2} 
\right) \nonumber \\ 
&& - \frac{1}{2 q} \int_0^\infty d p \, p^2 \, [n_+ (p) + n_- 
(p)] \, 
\ln \left( \frac{L_{+ +} \, L_{- -}}{L_{+ -} \, L_{- +}} \right) 
,  \nonumber \\ 
\\ 
I_3 & \equiv & \frac{1}{2 q} \int_0^\infty d p \, p \, [n_+ (p) + 
n_- (p)] 
\, \ln \left( \frac{L_{+ +} \, L_{+ -}}{L_{- +} \, L_{- -}} \right) 
, \nonumber \\ 
&& \\ 
\tilde{I}_1 & \equiv & - \frac{1}{2 q} \int_0^\infty d p \, p \, 
[n_+ (p) - 
n_- (p)] \, \ln \left( \frac{L_{+ +} \, L_{- -}}{L_{+ -} \, L_{- 
+}} \right) , \nonumber \\ 
&& \\ 
\tilde{I}_3 & \equiv & \frac{1}{2 q} \int_0^\infty d p \, [n_+ (p) - 
n_- (p)] 
\, \ln \left( \frac{L_{+ +} \, L_{+ -}}{L_{- +} \, L_{- -}} \right) 
\label{full} 
\end{eqnarray} 
with 
\begin{eqnarray*} 
n_\pm (p) & = & 1 / (e^{(p \mp \mu)/ T} + 1) , \\ 
L_{\rho \sigma} & \equiv & q_0 + \rho q + 2 \sigma p \;\;\;\;\; 
(\rho, \sigma = \pm) . 
\end{eqnarray*} 

It is straightforward to obtain 
\begin{eqnarray}
\mbox{Im} \Pi_F^{(T / L)} (Q) & \equiv &  \frac{1}{2 i} \left[ 
\Pi_F^{(T / L)} (q_0 (1 + i \epsilon), q) - \mbox{c.c.} \right] , 
\nonumber \\ 
\label{BA} \\ 
\mbox{Im} F_T (Q) & = & \theta (- Q^2) \, \frac{3 \pi}{4} \, 
m_\gamma^2 \, Q^2 \, \frac{|q_0|}{q^3} , 
\label{15} \\ 
\mbox{Im} \left( \frac{F_L (Q)}{Q^2} \right) & = & - \theta (- Q^2) 
\, \frac{3 \pi}{2} \, m_\gamma^2 \, \frac{|q_0|}{q^3} , 
\label{Im-pi} \\ 
\mbox{Im} \, I_1 & = & - \frac{\pi T}{2 q} \left( F_{- -} + F_{+ -} 
- F_{- +} - F_{+ +} \right) , \nonumber \\ 
&& \\ 
\mbox{Im} \, I_2 & = & - \frac{\pi}{2 q} \int_{q_l}^{q_u} d p \, p^2 
[ n_+ (p) + n_- (p) ] , \\ 
\mbox{Im} \, I_3 & = & - \frac{\pi}{2 q} \epsilon (q_0) 
\int_{q_l}^{q_u} d p \, p [ n_+ (p) + n_- (p) ] \nonumber \\ 
& & - \frac{\pi}{q} \epsilon (q_0) \theta (- Q^2) \nonumber \\ 
&& \times \int_0^{q_l} d p 
\, p \, [ n_+ (p) + n_- (p) ] , \\ 
\mbox{Im} \tilde{I}_1 & = & - \frac{\pi}{2 q} \int_{q_l}^{q_u} d p 
\, p \, \left[ n_+ (p) - n_- (p) \right] , \\ 
\mbox{Im} \tilde{I}_3 & = & \frac{\pi}{q} \epsilon (q_0) \theta (- 
Q^2) \mu \nonumber \\ 
&& - \frac{\pi}{2 q} \epsilon (q_0) T \left[ F_{+ +} - F_{- +} 
\right. \nonumber \\ 
&& \left. - \epsilon (Q^2) \left( F_{+ -} - F_{- -} \right) \right] 
, 
\label{iti} 
\end{eqnarray}
where 
\begin{eqnarray} 
q_u & \equiv & \frac{|q_0| + q}{2} \, , \;\;\;\;\;\;\; q_l \equiv 
\frac{||q_0| - q|}{2} , \nonumber \\ 
F_{\rho \sigma} & = & \ln \left( e^{\rho \mu / T} + e^{- ||q_0| + 
\sigma q| / (2 T)} \right) \;\;\;\;\;\;\; (\rho, \sigma = + , -) 
. \nonumber \\ 
\end{eqnarray} 
Note that $\Pi_F^{(V \!A)} (Q)$ vanishes for $\mu = 0$. 
\end{appendix} 
 
\begin{figure} 
\caption{Diagram for $\tilde{\Gamma}_d^{(1)}$. \lq\lq $1$'' (\lq\lq 
$2$'') at the vertex on the left-side (right-side) denotes the type 
of vertex in real-time equilibrium/nonequilibrium quantum field 
theory. The oval loop is an electron loop. 
\label{fig1} } 
\caption{
Integration region ${\cal R} = {\cal R}_1 \cup {\cal R}_2$. The 
dashed line with $T$ ($L$) shows the dispersion relation for the 
transverse (longitudinal) mode in the hard-thermal-loop resummed 
photon propagator. 
\label{fig2} } 
\caption{
Diagram for $\tilde{\Gamma}_d^{(2)}$. The hard-thermal-loop resummed 
effective photon propagator is indicated by a blob. \lq\lq $1$'', 
\lq\lq $2$'', \lq\lq $i$'', and \lq\lq $j$'' on the vertexes denote 
the type of vertex. The oval loops are electron loops. 
\label{fig3} } 
\caption{Plots of ${\cal G}$ and ${\cal G}^{(1)}$ vs $q_0$ at $T = 
50$ MeV, $\mu = 350$ MeV, and $E = 2$ MeV. Figure (a) corresponds to 
the processes (\ref{process1}) and (\ref{pro1}), and Fig.~(b) 
corresponds to the processes (\ref{process2}) and (\ref{pro2}). 
\label{fig4} } 
\caption{Same as in Fig.~4 but for $E = 10$ MeV. 
\label{fig5} } 
\caption{Same as in Fig.~4 but for $E = 20$ MeV. 
\label{fig6} } 
\caption{Same as in Fig.~4 but for $E = 50$ MeV. 
\label{fig7} } 
\caption{Same as in Fig.~4 but for $T = 20$ MeV and $E = 10$ MeV. 
\label{fig8} } 
\caption{Same as in Fig.~4 but for $\mu = 0$ and $E = 10$ MeV. 
\label{fig9} } 
\caption{Plots of $\tilde{\Gamma}_d$ and $\tilde{\Gamma}_d^{(1)}$ vs 
$E$ at $T = 50$ MeV and $\mu = 350$ MeV. 
\label{fig10} } 
\caption{Same as in Fig.~10 but for $T = 20$ MeV. 
\label{fig11} } 
\caption{Plots of $\tilde{\Gamma}_d$ and $\tilde{\Gamma}_d^{(1)}$ vs 
$E$ at $T = 50$ MeV and $\mu = 0$. 
\label{fig12} } 
\caption{Same as in Fig.~12 but for $T = 20$. 
\label{fig13} } 
\end{figure} 
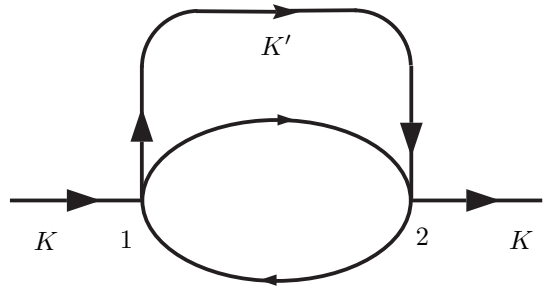
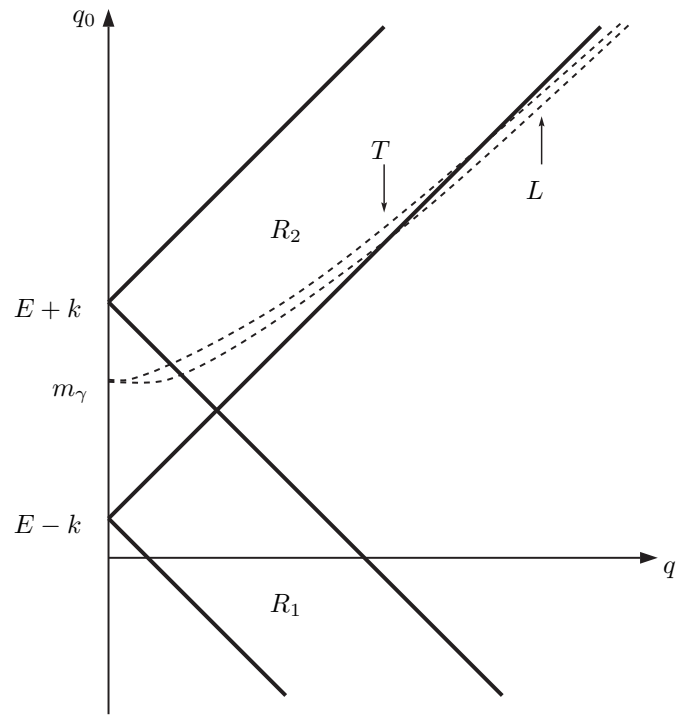
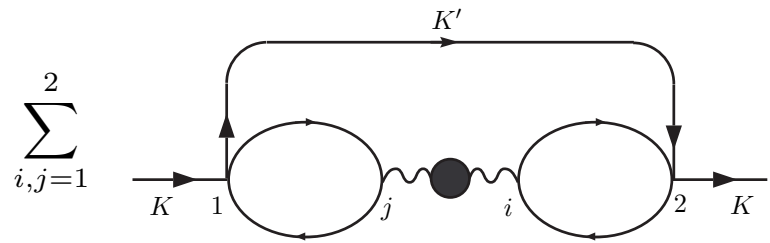
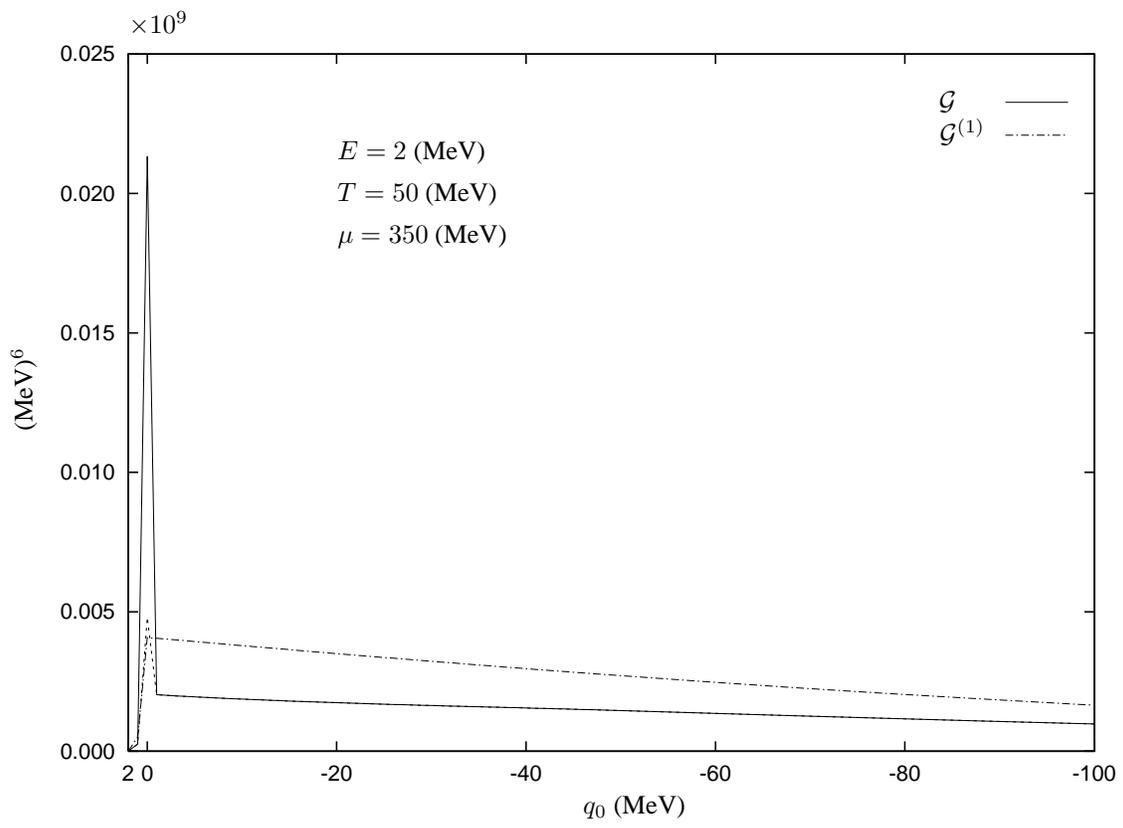
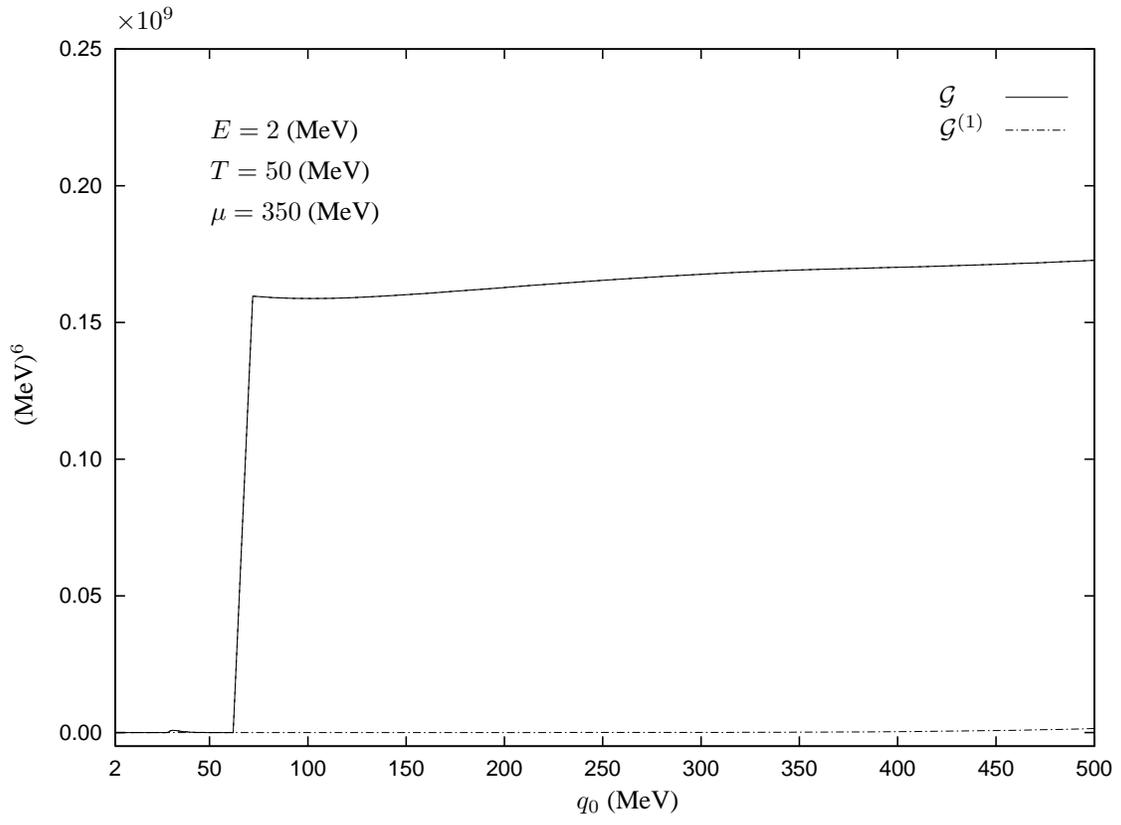
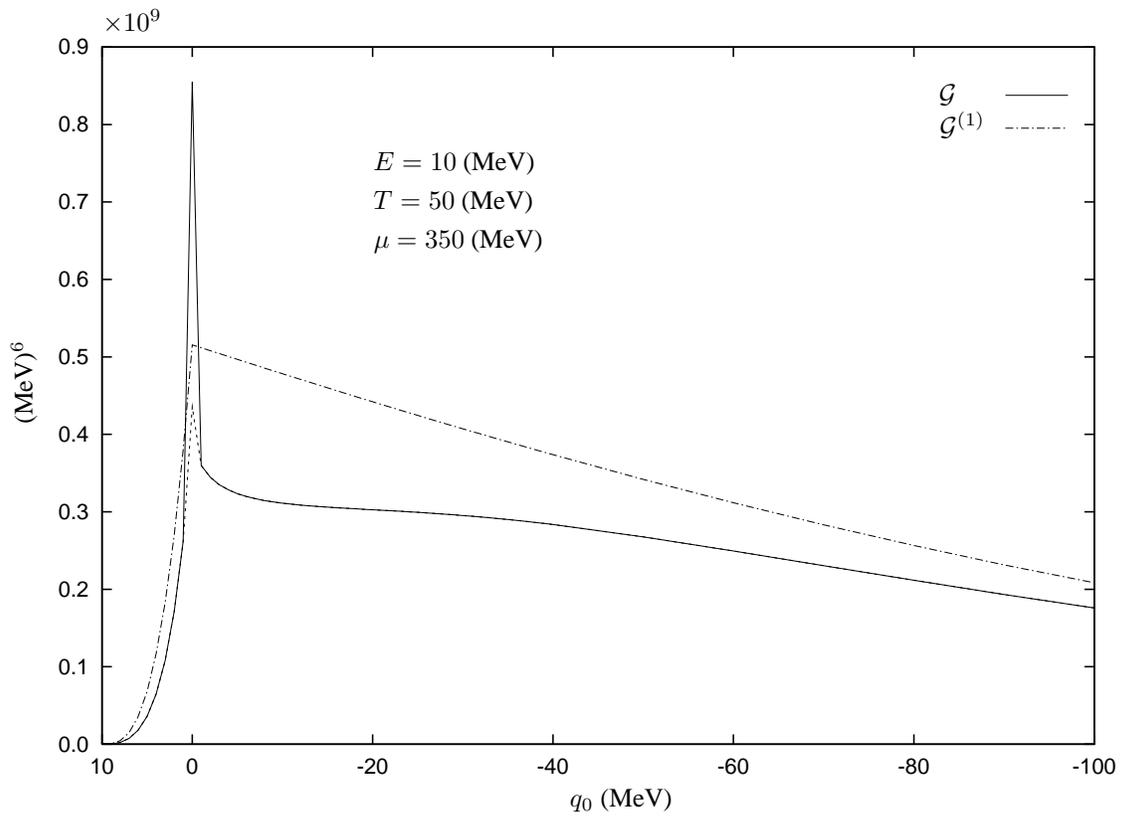
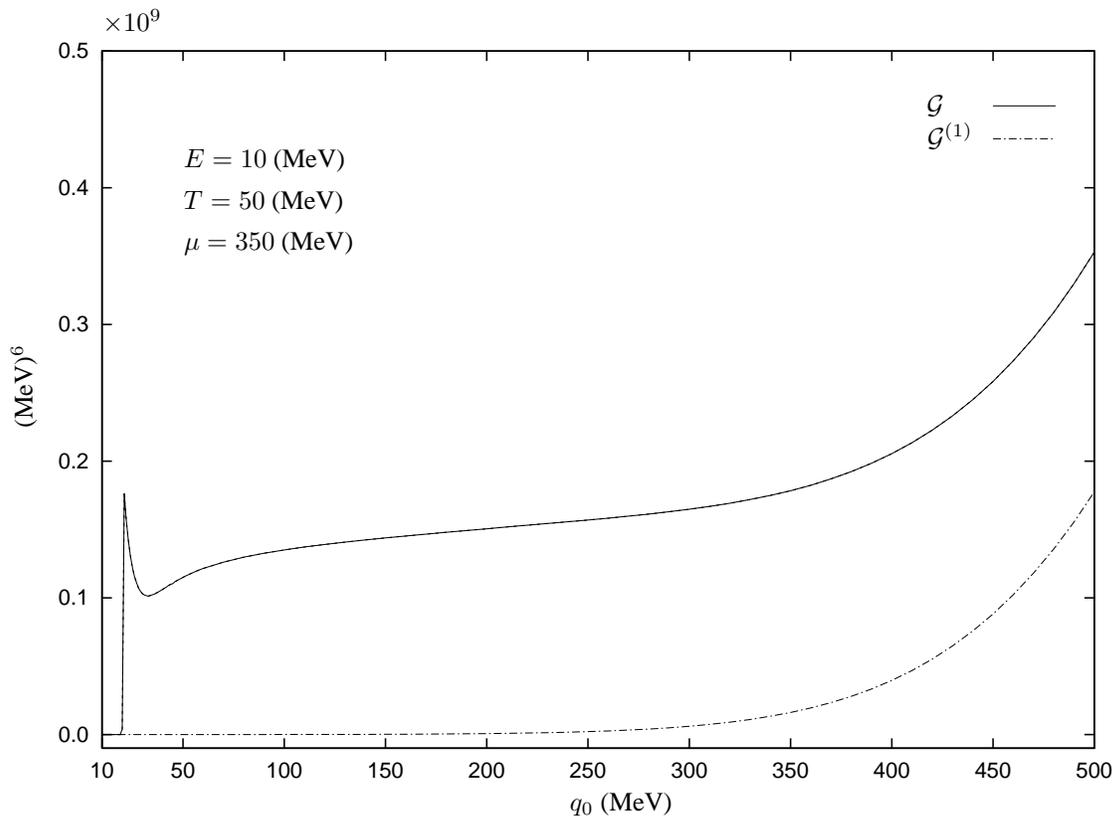
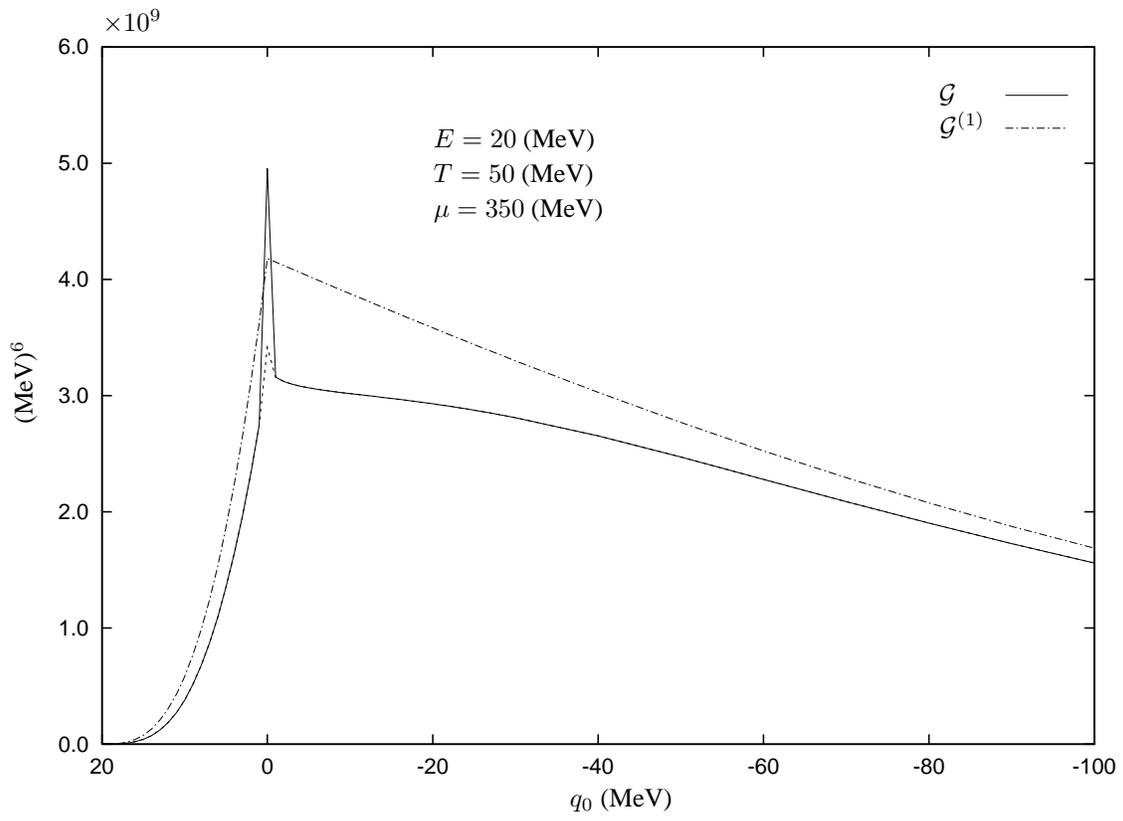
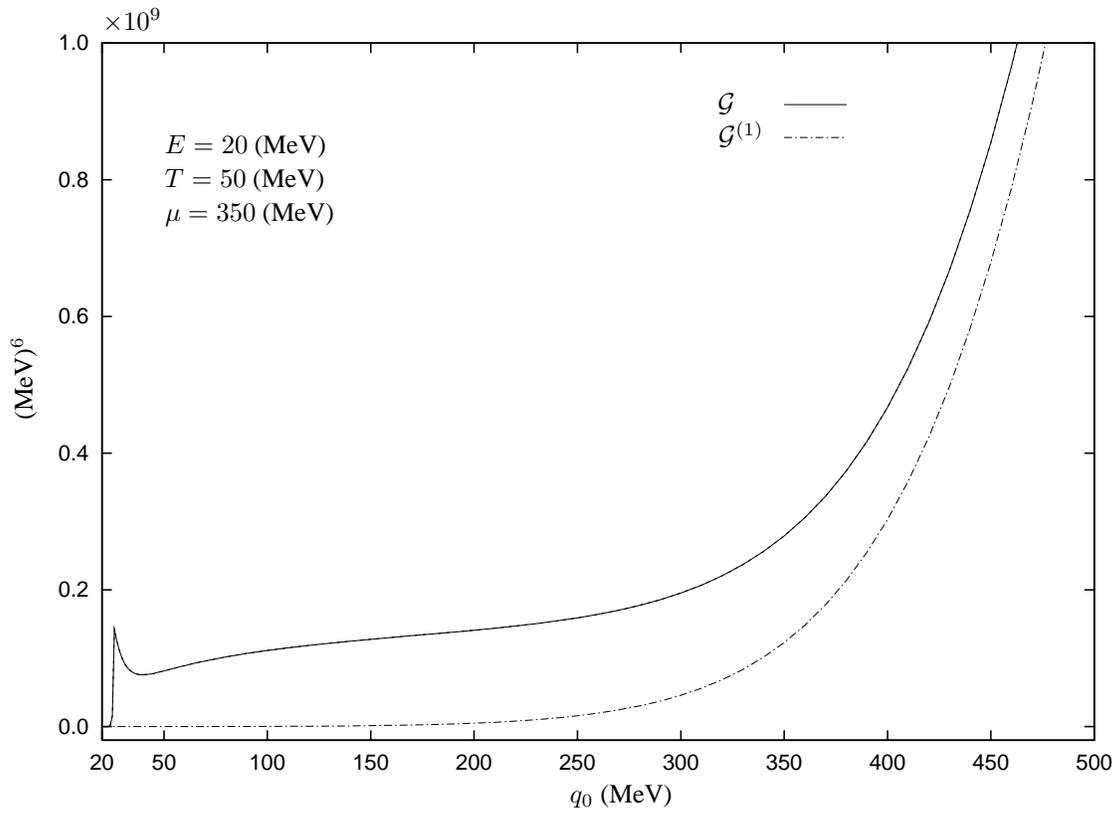
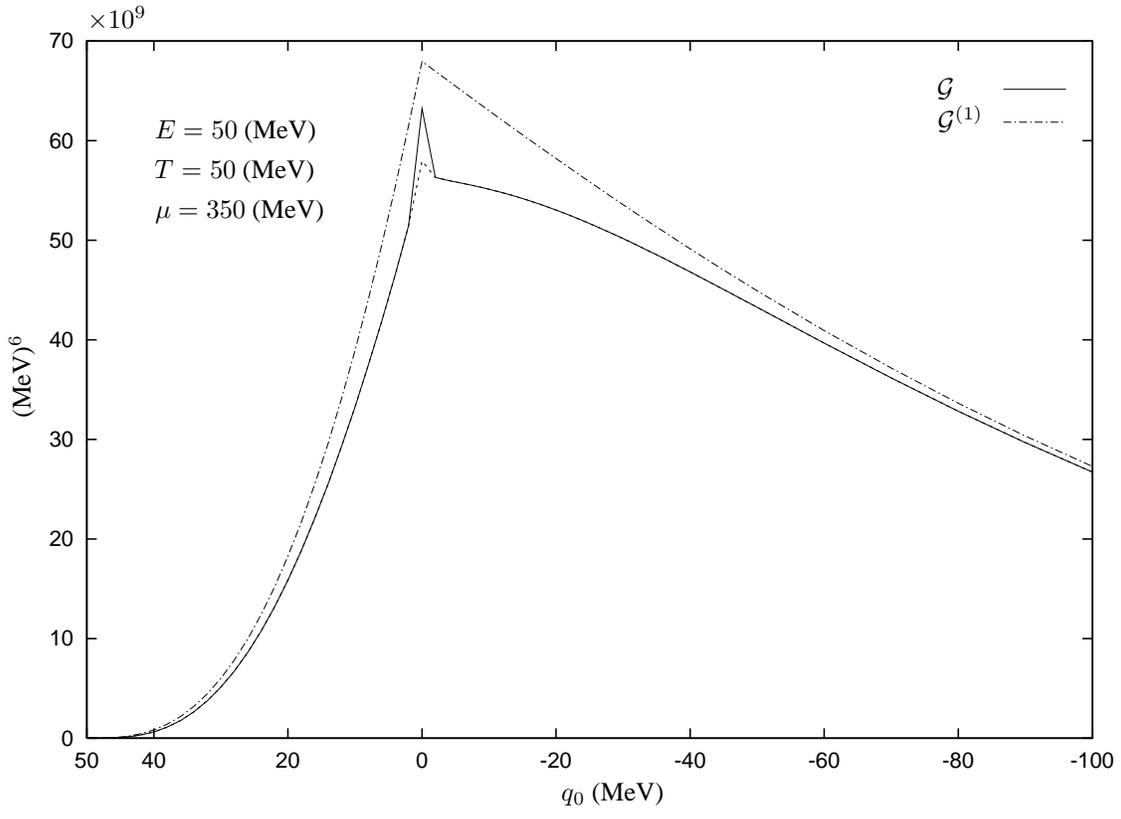
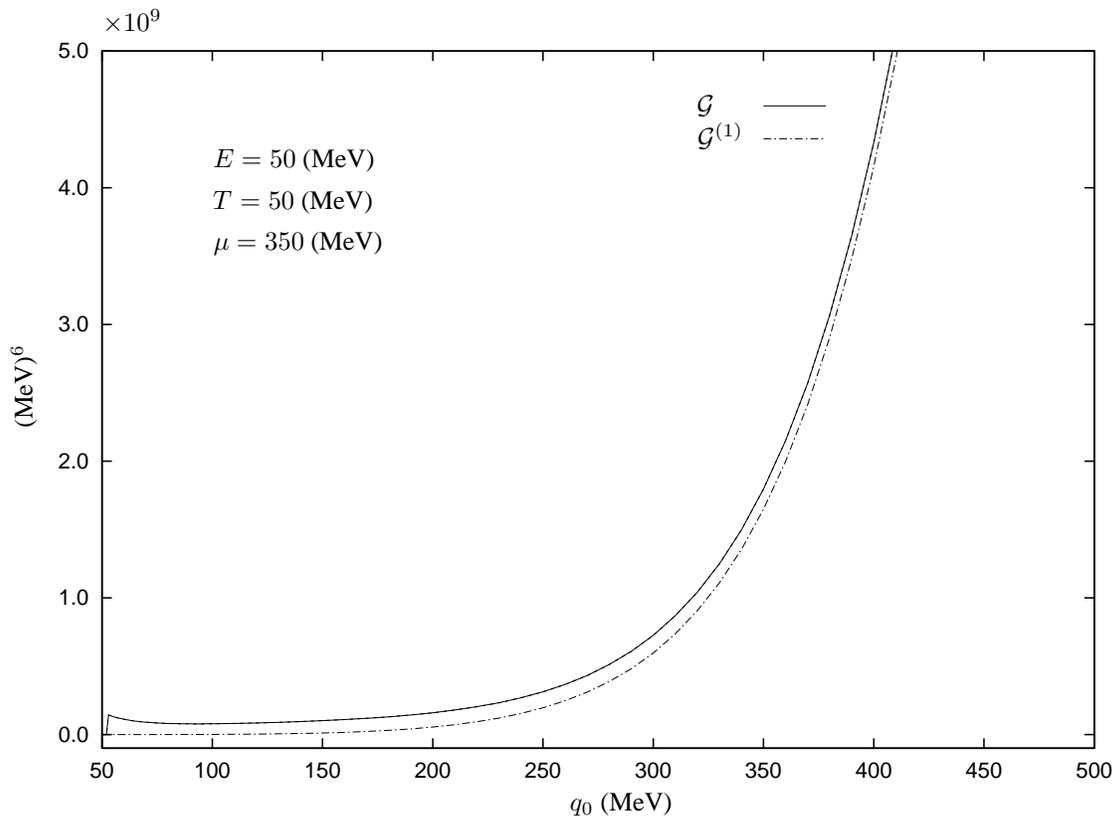
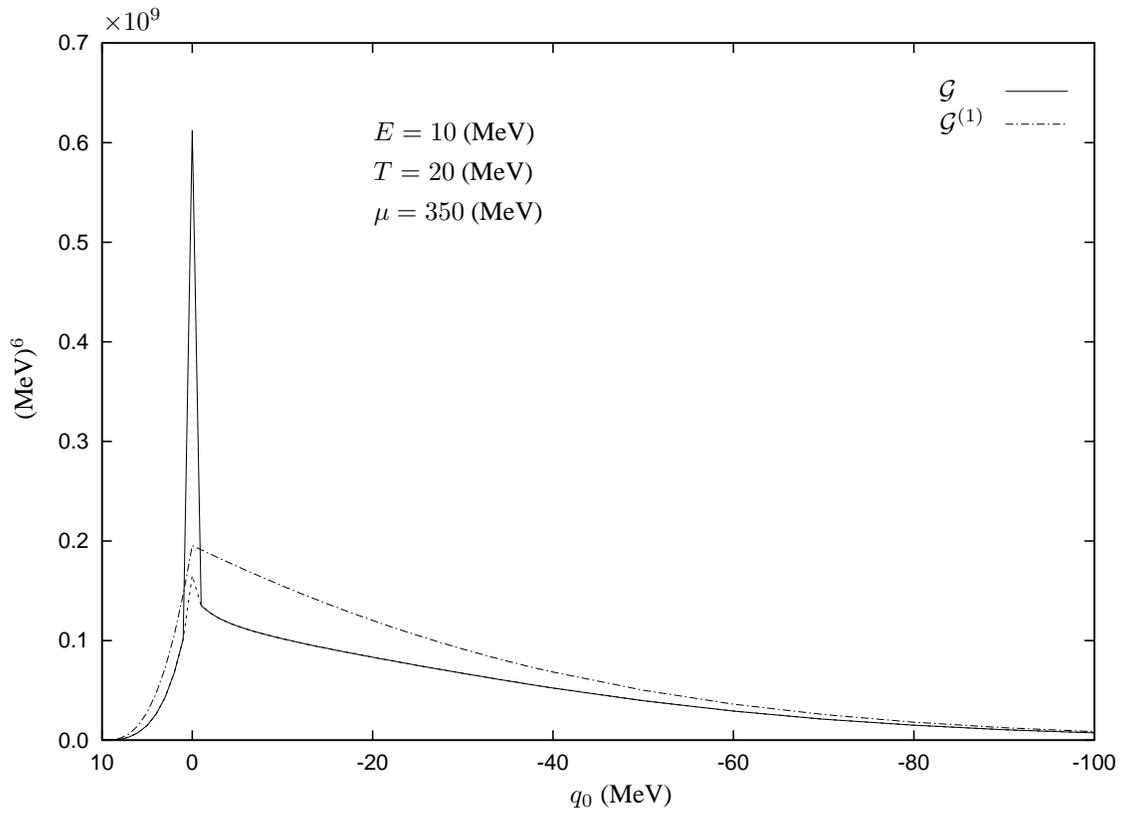
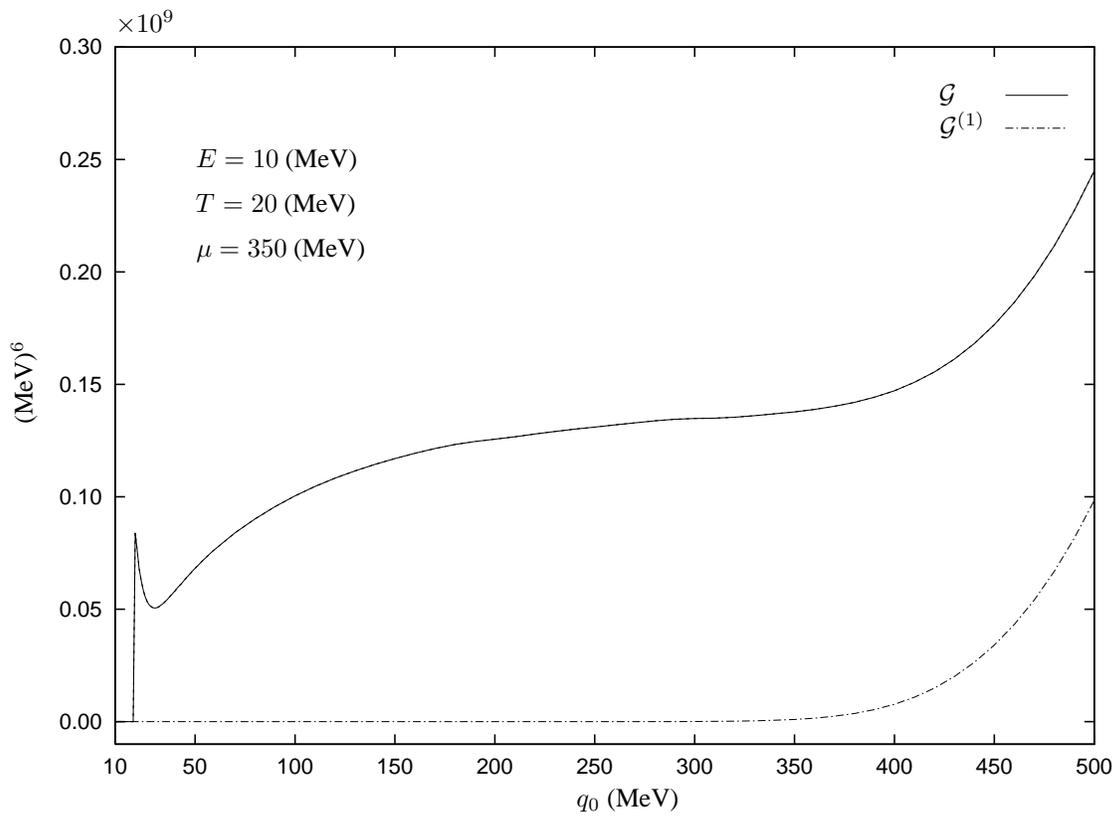
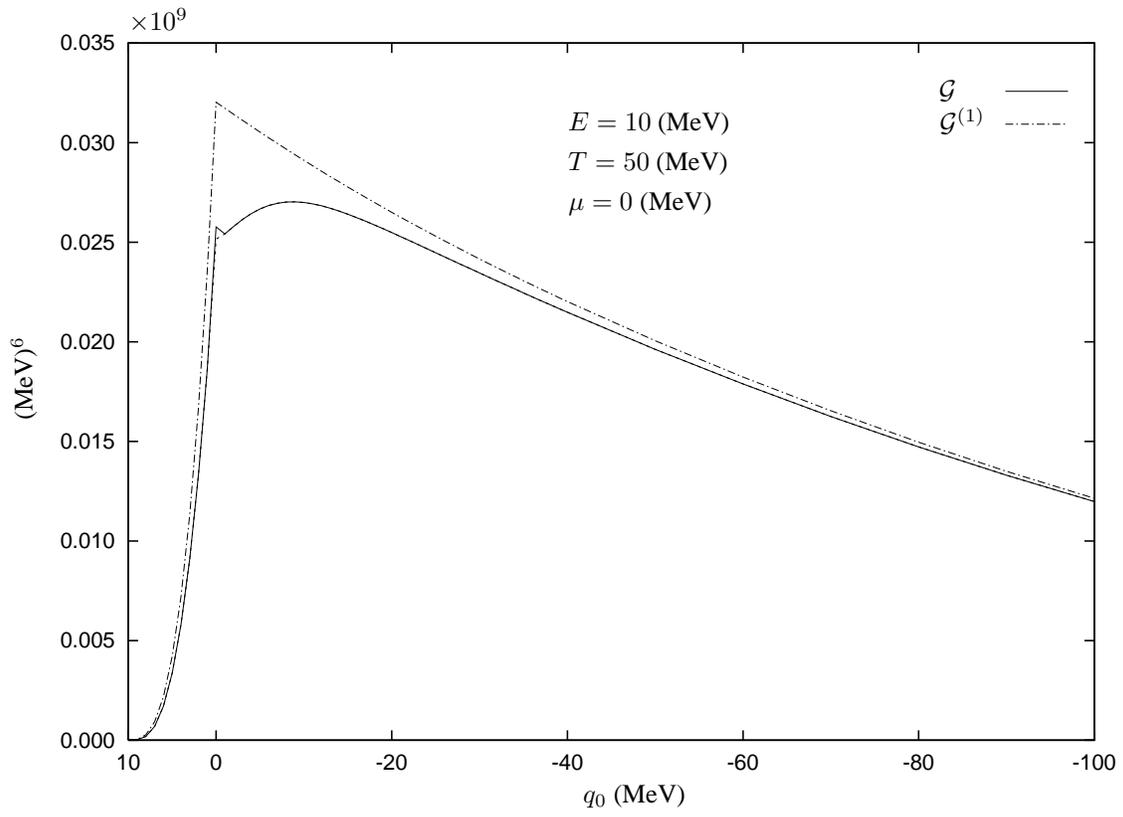
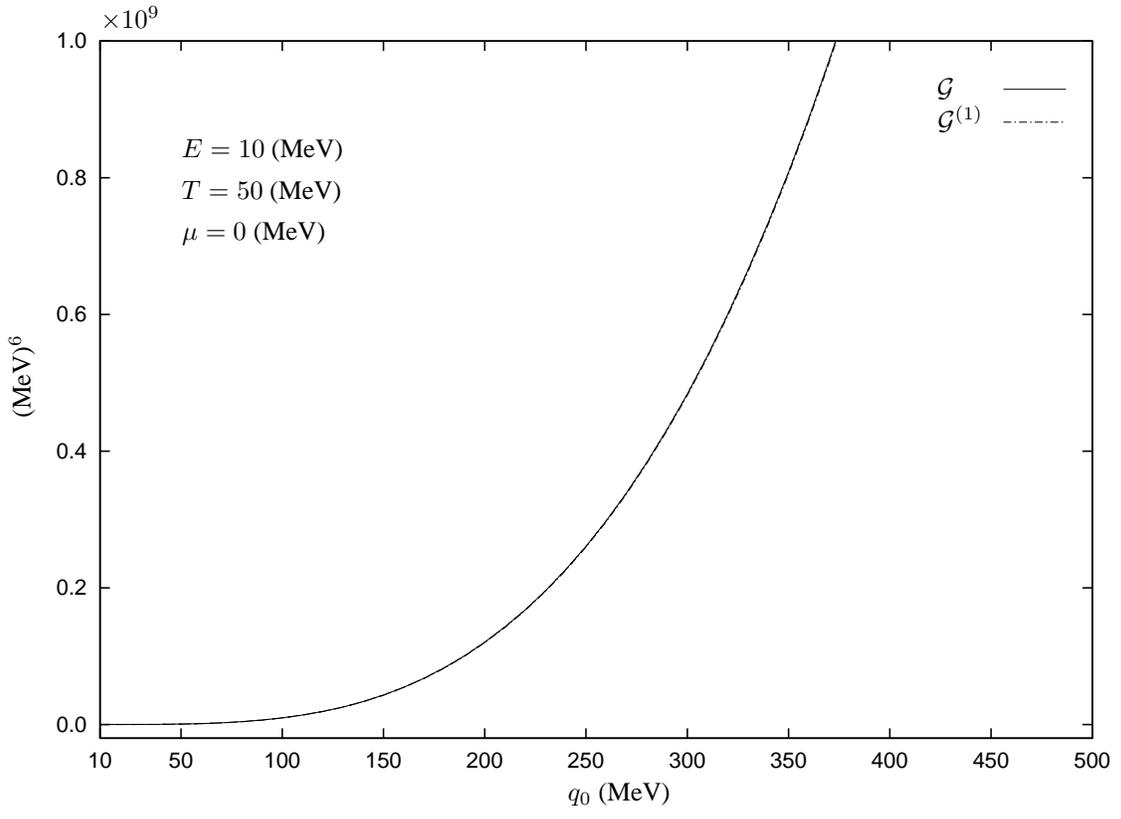
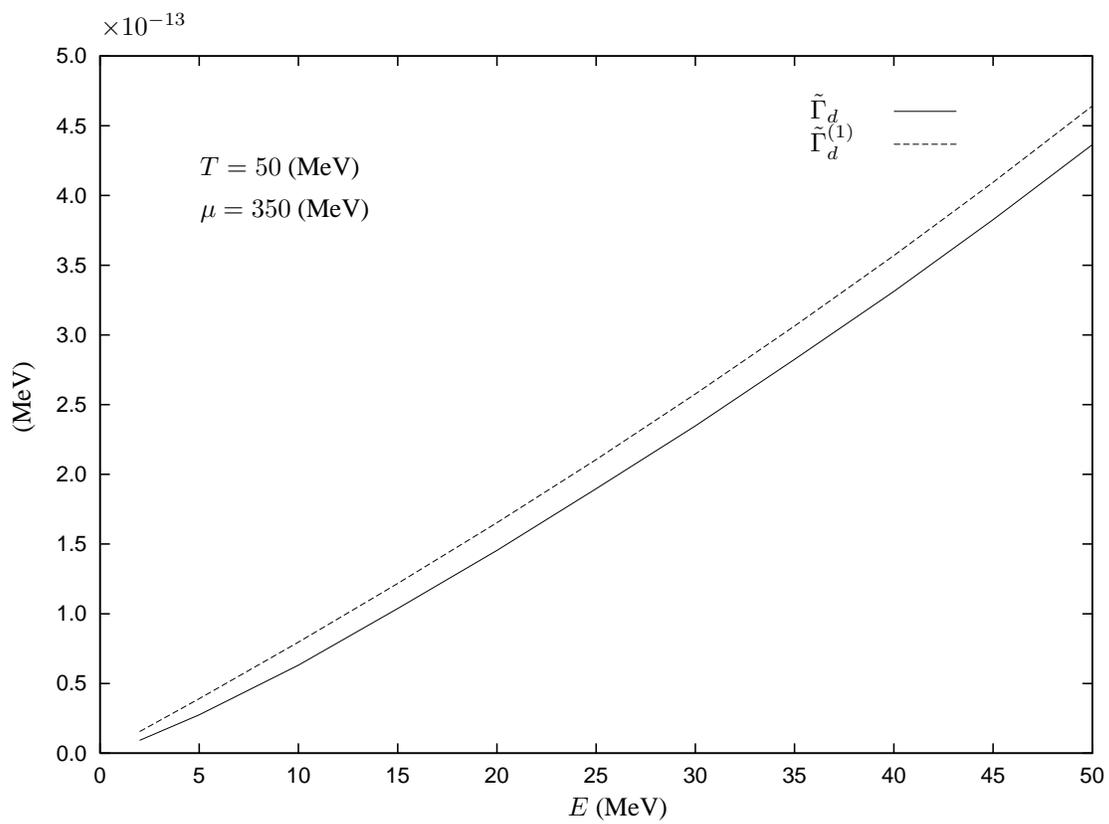
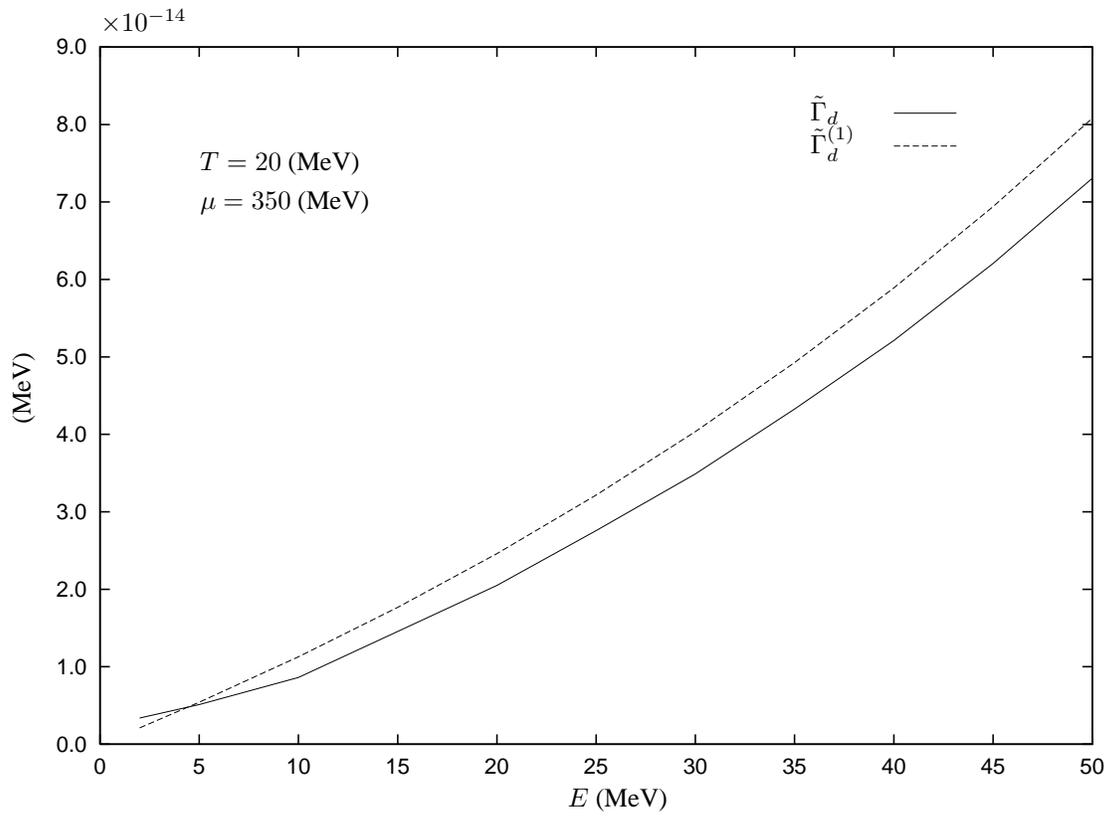
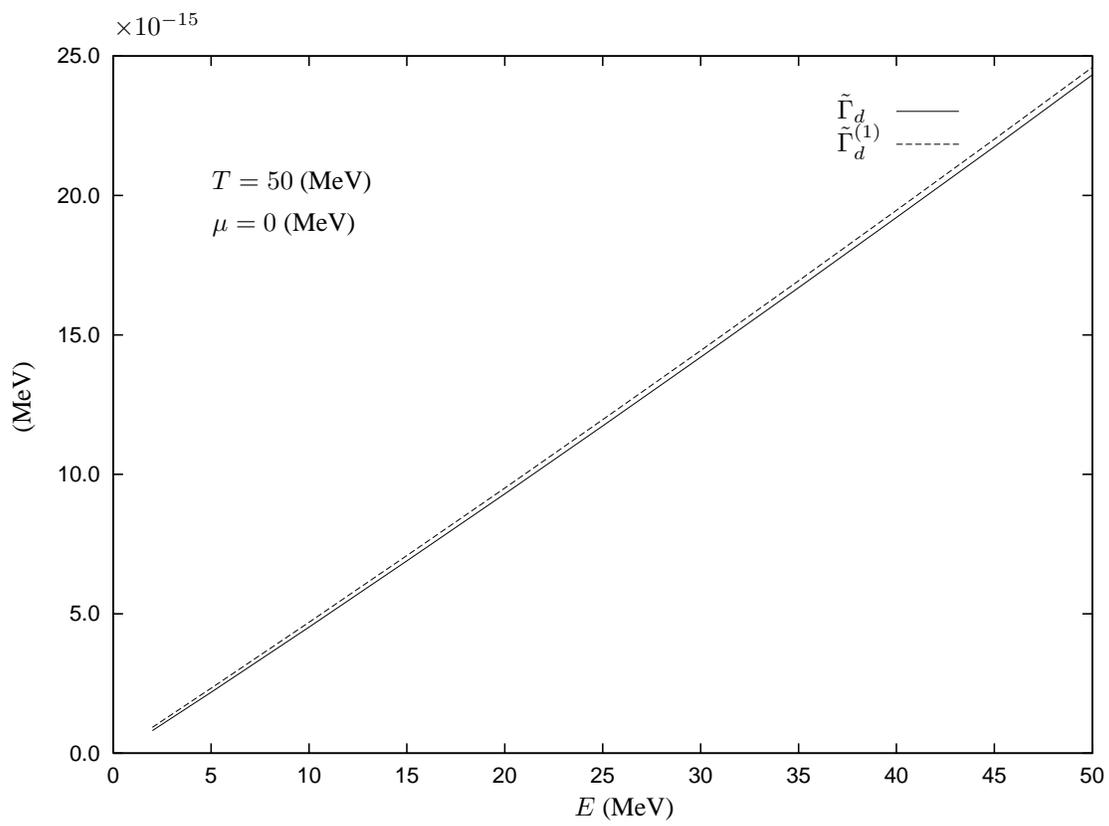
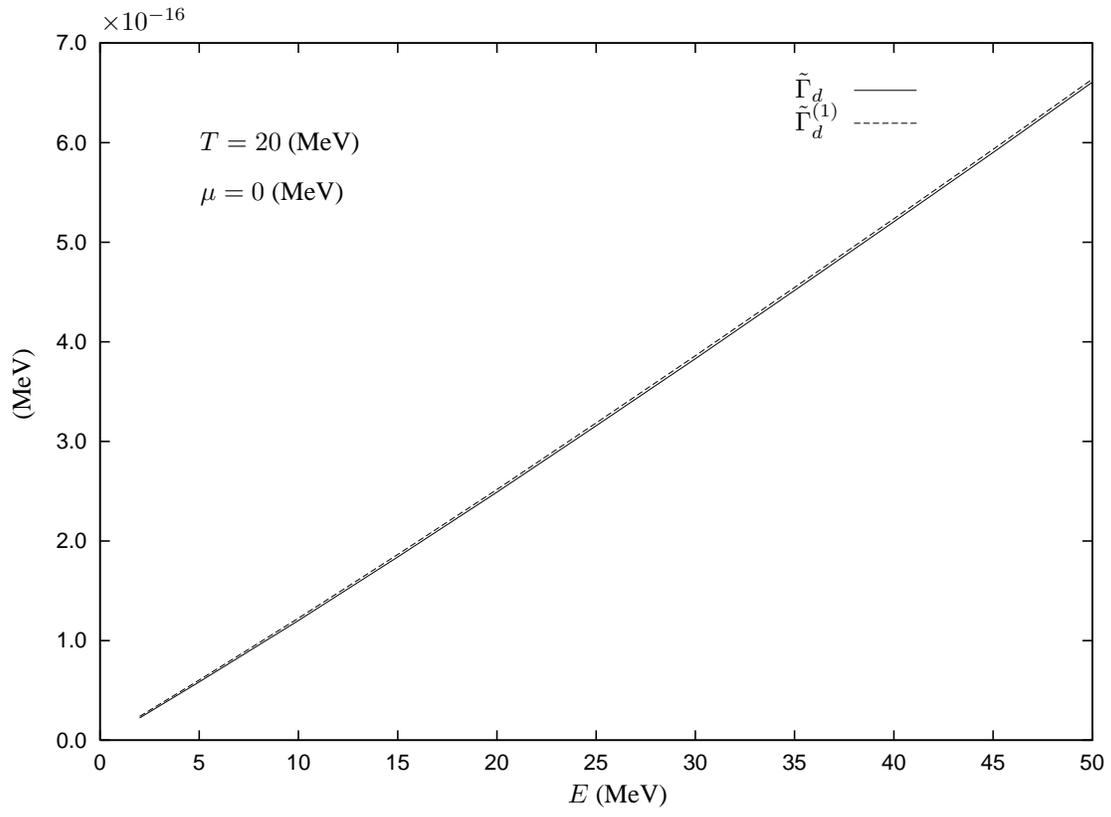
\end{document}